\documentclass[aps,prd, preprint, nofootinbib,superscriptaddress]{revtex4-1}
\usepackage{color}
\usepackage{graphicx}
\usepackage{ulem}
\begin{document}
\preprint{YITP-18-42}
\preprint{RIKEN-QHP-370}
\preprint{RIKEN-iTHEMS-Report-18}

\title{
Systematics of the HAL QCD Potential\\ at Low Energies  in  Lattice QCD}

\newcommand{\Tsukuba}{
Center for Computational Sciences, University of Tsukuba, Tsukuba 305-8577, Japan
}

\newcommand{\Kyoto}{
Center for Gravitational Physics, 
Yukawa Institute for Theoretical Physics, Kyoto University, Kitashirakawa Oiwakecho, Sakyo-ku, 
Kyoto 606-8502, Japan}

\newcommand{\Riken}{RIKEN Nishina Center, Wako 351-0198, Japan}

\newcommand{\RikenB}{RIKEN Interdisciplinary Theoretical and Mathematical Sciences Program (iTHEMS), Wako 351-0198, Japan}

\newcommand{\Nihon}{Nihon University, College of Bioresource Sciences, Kanagawa 252-0880, Japan}

\newcommand{\RCNP}{Research Center for Nuclear Physics (RCNP), Osaka University, Osaka 567-0047, Japan}

\author{Takumi~Iritani}
\affiliation{\Riken}

\author{Sinya~Aoki}
\affiliation{\Kyoto}
\affiliation{\Tsukuba}

\author{Takumi~Doi}
\affiliation{\Riken}
\affiliation{\RikenB}

\author{Shinya~Gongyo}
\affiliation{\Riken}

\author{Tetsuo~Hatsuda}
\affiliation{\RikenB}
\affiliation{\Riken}

\author{Yoichi~Ikeda}
\affiliation{\RCNP}

\author{Takashi~Inoue}
\affiliation{\Nihon}
 
\author{Noriyoshi~Ishii}
\affiliation{\RCNP}

\author{Hidekatsu~Nemura}
\affiliation{\RCNP}

\author{Kenji~Sasaki}
\affiliation{\Kyoto}

\collaboration{HAL QCD Collaboration}
\begin{center}
\end{center}

\begin{abstract}
The $\Xi\Xi$ interaction in the $^1$S$_0$ channel  is studied 
to examine the convergence of the derivative expansion of the non-local HAL QCD potential  at  the next-to-next-to-leading order (N$^2$LO).
We find that
(i) the leading order potential from the N$^2$LO analysis gives the scattering phase shifts accurately at low energies,
(ii)  the full N$^2$LO potential gives only small correction to the phase shifts
even at higher energies below the inelastic threshold, and 
(iii) the potential determined from the wall quark source at the leading order analysis 
agrees with the one at the N$^2$LO analysis except at short distances, and thus, it gives
correct phase shifts at low energies. 
We also study the possible systematic uncertainties in the HAL QCD potential
such as the inelastic state contaminations
and the finite volume artifact for the potential
and find that they  are well under control for this particular system.
\end{abstract}

%\begin{document}

\maketitle

%%%%%%%%%%%%%%%%%%%%%%%%%%%%%%%%%%%%%%%%%%%%%%%%%%%%%%%%%%%
\section{Introduction}
\label{sec:introduction}

In lattice QCD, two methods have been proposed so far to study the baryon-baryon interactions.
One is the direct method~\cite{Yamazaki:2015asa,Wagman:2017tmp, Berkowitz:2015eaa},
where the energy spectrum on finite volume(s) is extracted from the temporal correlation of two baryons
and is converted to the scattering phase shift and/or the binding energy in the infinite volume through 
the L\"uscher's finite volume formula~\cite{Luscher:1985dn,Luscher:1990ux}.
The other is the HAL QCD method~\cite{Ishii:2006ec,Aoki:2009ji,HALQCD:2012aa,Aoki:2012tk,Aoki:2012bb},
where the potential between baryons is first derived from the spatial correlations of two baryons,
and it is used to calculate the observables  through
the Schr\"odinger-type equation in the infinite volume.

While both methods are supposed to give  the same results in principle,
previous numerical studies for two-nucleon ($NN$) systems show clear discrepancy:
The direct method indicates that both dineutron ($^1$S$_0)$ and deuteron ($^3$S$_1$) 
are bound for heavy pion masses ($m_{\pi} \geq 300$~MeV),
while the HAL QCD method does not provide such bound states in both  channels for heavy pion masses.
This discrepancy was recently discussed in 
 a series of papers~\cite{Iritani:2016jie,Iritani:2017rlk,Aoki:2017byw,Iritani:2018vfn}, 
where it was pointed out that the effective two-particle energy as a function of the 
Euclidean time may significantly suffer from elastic scattering states of two nucleons.
To elucidate such  uncertainties, certain ``normality checks''
 for the finite-volume spectrums were introduced~\cite{Iritani:2016jie,Iritani:2017rlk,Aoki:2017byw,Iritani:2018vfn}.

The advantage of the time-dependent HAL QCD method~\cite{HALQCD:2012aa}
over the direct method is that
the former is free from the ground state saturation problem in principle,
since the energy-independent potential  controls
both ground state and the elastic excited states simultaneously
as long as the inelastic scatterings in the small Euclidean time are properly suppressed.\footnote{
Otherwise, the coupled channel  HAL QCD method should be used to take into account
the inelastic states \cite{Aoki:2012bb}.}
In practice,
there appear systematic uncertainties associated with the truncation of the derivative expansion
for the non-local potential.
Therefore, the main purpose of the present  paper is to study the convergence of the derivative expansion,
as well as other sources of systematic uncertainties such as
the inelastic state contaminations and the  distortion of the interaction under finite volume.
We consider the $\Xi\Xi$ system in the $^1$S$_0$ channel
and perform the (2+1)-flavor lattice QCD calculation at $m_\pi = 0.51$~GeV and $m_K = 0.62$~GeV.
    Because of the large quark masses,  the statistical errors in this case become relatively  small, so that one can focus on 
    the detailed analysis of the systematic errors.  Also,     this channel  and the $NN$ system  in the $^1$S$_0$ channel belong 
   to the same multiplet in the flavor SU(3) limit. 

This paper is organized as follows.
In Sec.~II, we review the time-dependent HAL QCD method.
In Sec.~III, we present the lattice QCD results for the $\Xi\Xi$ interaction in the $^1$S$_0$ channel  
at the next-to-next-to-leading order (N$^2$LO) in the derivative expansion.
The N$^2$LO potential is extracted from a specific combination of the  $\Xi\Xi$ correlations 
 with different source operators. The systematic errors associated with
the inelastic state contaminations and the distortion in the finite volume
are also examined.
In Sec.~IV, we calculate the scattering phase shifts in this channel,
and check the convergence of the derivative expansion in the HAL QCD method.
In Sec.~V,
we  demonstrate the self-consistency between the phase shifts obtained from the HAL QCD potential and
    those  obtained from the energy spectra  obtained from the HAL QCD potential combined with
  the  L\"uscher's  formula.  Sec.~VI is devoted to the conclusion. 
  In Appendix A, we discuss the relation between the energy-independent non-local potential
    and the energy-dependent local one.

\section{Formalism}
The key quantity in the HAL QCD method~\cite{Ishii:2006ec,Aoki:2009ji,HALQCD:2012aa,Aoki:2012tk,Aoki:2012bb}
 is the Nambu-Bethe-Salpeter (NBS) wave function, defined by
\begin{eqnarray}
\psi^W(\vec{r}) &=& \sum_{\vec{x}} \langle 0 \vert T\{ B(\vec{x}+\vec{r},0)B(\vec{x},0) \} \vert 2B, W \rangle,
\label{eq:NBS-def}
\end{eqnarray}
where $\vert 0 \rangle $ is the vacuum state of QCD, $\vert 2B, W\rangle$ is the QCD eigenstate  for two baryons with eigenenergy $W$, and $B(\vec{x},t)$ is a single baryon operator
with spin indices omitted for simplicity.
We then define  a non-local and  energy-independent potential  $ U(\vec{r}, \vec{r'})$
so as to satisfy 
\begin{equation}
  (E_k - H_0) \psi^W(\vec{r})
  =\int d\vec{r'}\  U(\vec{r}, \vec{r'})
  \psi^W(\vec{r'})
  \label{eq:t-indep-HAL}
\end{equation}
below inelastic threshold, $W < W_\mathrm{th} = 2 m_B + m_\pi$, with $m_B$  the baryon mass, $m_\pi$ the pion mass,
 and $W = 2\sqrt{m_B^2 + k^2}$.
  Here we define $E_k = k^2/(2\mu)$ and $H_0 = -\nabla^2/(2\mu)$ with a reduced mass $\mu = m_B/2$. 
We note that $U(\vec{r}, \vec{r'})$ depends on the specific choice of the interpolating operator $B(x)$ used in
Eq.~(\ref{eq:NBS-def}).  Nevertheless, the $S$-matrix  is free from the  choice of $B(x)$ 
 as long as  it is an ``almost-local operator field''  \cite{Haag:1958vt}  (Nishijima-Zimmermann-Haag theorem).

To extract the NBS wave function in lattice QCD, 
we start with the two-baryon correlation function,
\begin{equation}
  C_{2B}(\vec{r}, t-t_0) =  \sum_{\vec{x}} \langle 0 | T \{ B(\vec{x}+\vec{r},t)B(\vec{x},t) 
  \overline{\mathcal{J}}_{2B}(t_0) \}| 0\rangle,
\end{equation}
where $\overline{\mathcal{J}}_{2B}(t_0)$ is a source operator for two-baryon.
By inserting the complete set, we obtain
\begin{eqnarray}
  C_{2B}(\vec{r},t-t_0) &=& \sum_{\vec{x}} \langle 0 | T\{ B(\vec{x}+\vec{r},t) B(\vec{x},t)\}
  \sum_{n} | 2B, W_n \rangle \langle 2B, W_n  |
  \overline{\mathcal{J}}_{2B}(t_0) | 0 \rangle + \cdots \nonumber \\
  &=& \sum_{n } A_{n } \psi^{W_n}  (\vec{r})e^{-W_n(t-t_0)} + \cdots,
\end{eqnarray}
where $W_n =2\sqrt{m_B^2+k_n^2}$ is the $n$-th energy eigenvalue,
$A_{n} \equiv \langle 2B, W_n |\overline{\mathcal{J}}_{2B}(0)|0\rangle$
corresponds to the overlap with each elastic eigenstate,
and the ellipses represent the  inelastic contributions.
In principle, one can extract $A_0 \psi^{W_0}(\vec{r})$ for the lowest energy $W_0$ from the large $t$ behavior of $C_{2B}(\vec{r},t)$.

In practice, however, since $C_{2B}(\vec{r},t)$ becomes too noisy at large $t$,
we need to employ the time-dependent HAL QCD method~\cite{HALQCD:2012aa}.
Let us define the ratio of correlation functions, which we call the $R$-correlator, as
\begin{equation}
  R(\vec{r}, t) \equiv \frac{C_{2B}(\vec{r}, t)}{\{C_B(t) \}^2}
  = \sum_{n} A'_n \psi^{W_n}(\vec{r}) e^{-\Delta W_n t} + \mathcal{O}(e^{- \Delta W_\mathrm{th}t})
  \label{eq:R-correlator}
\end{equation}
with $\Delta W_n = W_n - 2m_B$, 
$\Delta W_\mathrm{th} = W_\mathrm{th} - 2m_B$
  and
$A'_n = A_n / {\cal C}^2$,
where $C_B(t)$ and ${\cal C}$ are a single baryon correlation function and the corresponding overlap factor, respectively. 
They are given by
\begin{eqnarray}
  C_B (t-t_0) = \sum_{\vec{x}} \langle 0 | T\{ B(\vec{x},t) \overline{\mathcal{J}}_{B}(t_0)  \} | 0 \rangle
  = {\cal C} \cdot e^{-m_B (t-t_0)} + \cdots ,
\end{eqnarray}
where $\overline{\mathcal{J}}_{B}(t_0)$ is a single baryon source operator
and ellipses represent the inelastic states contributions.

Since the non-local potential $U(\vec{r}, \vec{r'})$ is defined to be 
energy-independent~\cite{Aoki:2009ji},
all elastic scattering states below the threshold share the same $U(\vec{r}, \vec{r'})$.
Therefore, Eq.~(\ref{eq:t-indep-HAL}) 
with an identity $\Delta W_n = {k_n^2}/{m_B} - {(\Delta W_n)^2}/{(4m_B)}$ leads to
\begin{equation}
    \left[
      - H_0
      - \frac{\partial}{\partial t} 
      + \frac{1}{4m_B} \frac{\partial^2}{\partial t^2}
    \right]R(\vec{r},t)
  = \int d\vec{r'} \ U(\vec{r}, \vec{r'}) R(\vec{r'},t),
 \label{eq:master-equation} 
\end{equation}
where the effect of the inelastic channel of $\mathcal{O}(e^{-\Delta W_\mathrm{th}t})$ is neglected
in the right hand side, while there is no term beyond $\partial^2/\partial t^2$ in the left hand side of
Eq.~(\ref{eq:master-equation}), i.e., Eq.~(\ref{eq:master-equation}) is derived without non-relativistic approximation.

Note that the ground state saturation is no more required in this time-dependent HAL QCD method.
Instead, the required condition is that
$R(\vec{r},t)$ is saturated by the contributions from elastic states (``the elastic state saturation''),
which can be achieved by a moderate value of $t$ 
($\sim {\cal O}({\rm min.}\{\Lambda_{\rm QCD}^{-1}, m_{\rm NG}^{-1}\})$ with
$m_{\rm NG}$ being the mass of the lightest Nambu-Goldstone boson).\footnote{
  There is a possibility that the inelastic contributions 
   cancel partially between the numerator and the denominator of  $R(\vec{r},t)$, so that
      the elastic state saturation in  $R(\vec{r},t)$ may appear for smaller $t$ than 
    those in $C_{2B}(\vec{r}, t)$ and $C_B(t)$.}
    This is the fundamental difference between the HAL QCD method and the direct method.

As discussed in \cite{Aoki:2012tk,Aoki:2012bb},
 $U(\vec{r}, \vec{r'})$ in Eq.~(\ref{eq:master-equation})  is not determined  uniquely   by $R(\vec{r},t)$,
 though different  $U(\vec{r}, \vec{r'})$s give same observables below the inelastic threshold.
  In the HAL QCD method, the derivative expansion scheme enables one to 
   extract one of the possible $U(\vec{r}, \vec{r'})$s in a unique manner.
Let us consider the two-baryon system in the spin-singlet channel. Then
 the leading order (LO)  analysis neglecting the higher orders leads to 
 \begin{equation}
 U(\vec{r}, \vec{r'}) 
  = V_0^\mathrm{LO} (r) \delta(\vec{r} - \vec{r'}),
   \label{eq:LO-expansion}
\end{equation}
with
\begin{equation}
  V_0^\mathrm{LO}(r) =
    - \frac{H_0 R(\vec{r},t)}{R(\vec{r},t)} 
    - \frac{(\partial/\partial t) R(\vec{r},t)}{R(\vec{r},t)}
    + \frac{1}{4m_B} \frac{(\partial^2/\partial t^2) R(\vec{r},t)}{R(\vec{r},t)} .
  \label{eq:veff}
\end{equation}
   In order to examine the convergence of the derivative expansion, 
 we consider the N$^2$LO analysis in this paper,
\begin{equation}
 U(\vec{r}, \vec{r'}) 
 = \{ V_0^\mathrm{N^2LO}(r) + V_2^\mathrm{N^2LO}(r)\nabla^2\}\delta(\vec{r} - \vec{r'}).
\label{eq:NLO-expansion}
\end{equation}
The relation between the potential from the LO analysis, $V_0^\mathrm{LO}(r)$,
and  those from the N$^2$LO analysis,
$V_0^\mathrm{N^2LO}(r)$ and $V_2^\mathrm{N^2LO}(r)$
is given by
\begin{equation}
  V_0^\mathrm{LO}(r)
  = V_0^\mathrm{N^2LO}(r) + V_2^\mathrm{N^2LO}(r) \frac{\nabla^2 R(\vec{r}, t)}{R(\vec{r},t)},  
  \label{eq:vnlo}
\end{equation}
which shows
  that the N$^2$LO correction in $V_0^\mathrm{LO}(r)$ depends on both $V_2^\mathrm{N^2LO}(r)$
  and the spatial profile of the $R$-correlator,
  the latter of which depends not only on
  the spatial profile of the NBS wave functions $\psi^{W_n}(r)$ but also on their magnitude $A_n^\prime $ in the $R$-correlator.
  The potentials $V_{0,2}^\mathrm{N^2LO}(r) $ 
   are $t$-independent as long as the elastic state saturation is achieved and the 
 higher order contributions in the derivative expansion can be neglected.
  One may also estimate the magnitude of systematic errors from the truncation of the derivative expansion
  and from the inelastic state contaminations by studying the $t$-dependence of
 the potentials.

\section{HAL QCD potential}

\subsection{Lattice Setup}

Throughout this paper, we use 2+1 flavor QCD ensembles~\cite{Yamazaki:2012hi},
generated by using  the Iwasaki gauge action and $\mathcal{O}(a)$-improved 
Wilson quark action at $a = 0.08995(40)$~fm
on $40^3 \times 48$, $48^3 \times 48$ and $64^3 \times 64$ lattice volumes
with heavy up/down quark masses and the physical strange quark mass,
$m_\pi = 0.51$~GeV, 
$m_K = 0.62$~GeV,
$m_N = 1.32$~GeV  
and $m_\Xi = 1.46$~GeV,
though only the one with the largest volume is used unless otherwise stated.
We employ the wall source 
$q^\mathrm{wall}(t) = \sum_{\vec{y}}q(\vec{y},t)$,
which has been mainly used in the previous studies by the HAL QCD method,
and the smeared source $q^\mathrm{smear}(\vec{x},t) = \sum_{\vec{y}} f(|\vec{x}-\vec{y}|)q(\vec{y},t)$
with the smearing function  $f(r) \equiv \{Ae^{-Br}, 1, 0\}$ for
\{$0 < r < (L-1)/2$, $r=0$, $(L-1)/2\leq r$\} ~\cite{Yamazaki:2012hi}.
 For the smeared source,
the same $\vec{x}$ is taken as the center of the smeared source for all  six quarks in two baryons
as has been done in Ref.~\cite{Yamazaki:2012hi}.
For both sources,
the point-sink operator for each baryon (``point-sink scheme'' in the HAL QCD method~\cite{Kawai:2017goq})
is exclusively employed in this study.
The correlation functions are calculated by the unified contraction algorithm (UCA)~\cite{Doi:2012xd}.
A number of configurations and other parameters are summarized in Table~\ref{tab:lattice_setup}.
Statistical errors are evaluated by the jack-knife method.
For more details on the simulation setup, see Ref.~\cite{Iritani:2016jie}.

In the present study, we focus on the $\Xi\Xi$ system in the $^1$S$_0$ channel:
This  is one of the most convenient choices to obtain the insights of $NN$ systems,
since it belongs to the same $\mathbf{27}$ representation
as the $NN$ system in the $^1$S$_0$ channel in the flavor SU(3) limit
but has much better signal to noise ratio than the $NN$($^1$S$_0$) case.
We use the relativistic interpolating operators~\cite{Iritani:2016jie} for $\Xi$,
which are given by
\begin{equation}
  \Xi_\alpha^0 = \varepsilon_{abc}(s^{aT}C\gamma_5 u^b)s_\alpha^c, \quad
  \Xi_\alpha^- = \varepsilon_{abc}(s^{aT}C\gamma_5 d^b)s_\alpha^c,
\end{equation}
where $C = \gamma_4\gamma_2$ is the charge conjugation matrix, 
$\alpha$ and ($a$, $b$, $c$) are the indices for the spinor and color, respectively.

\begin{table}
  \centering
  \begin{tabular}{|c|c|c|cc|c|}
    \hline
    \hline
    volume & $La$\ [fm] & \# of conf. & \# of smeared sources & $(A,B)$ & \# of wall sources \\
    \hline
    $40^3 \times 48$ & 3.6  & 207 & 512 & (0.8, 0.22) & 48 \\ 
    $48^3 \times 48$ & 4.3 & 200 & $4 \times 384$  & (0.8, 0.23) & $4 \times 48$ \\ 
    $64^3 \times 64$ & 5.8 & 327 & $1 \times 256$  & (0.8, 0.23) & $4 \times 64$ \\ 
    \hline
    \hline
  \end{tabular}
  \caption{Simulation parameters.
  The rotational symmetry for isotropic lattices is used to increase statistics.}
 \label{tab:lattice_setup}
\end{table}

\subsection{The $R$-correlator}

\begin{figure}
  \centering
  \includegraphics[width=0.49\textwidth,clip]{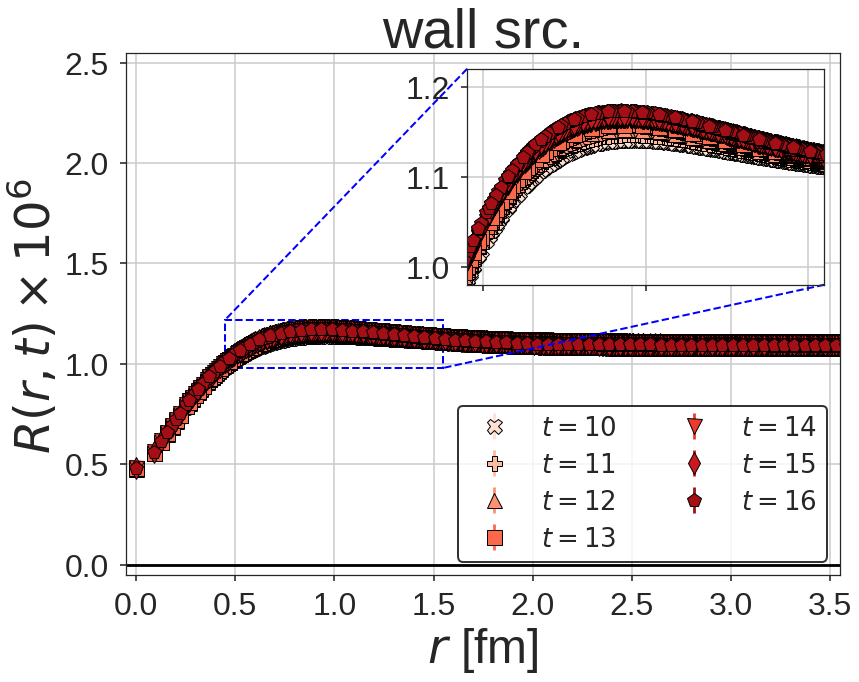}
  \includegraphics[width=0.49\textwidth,clip]{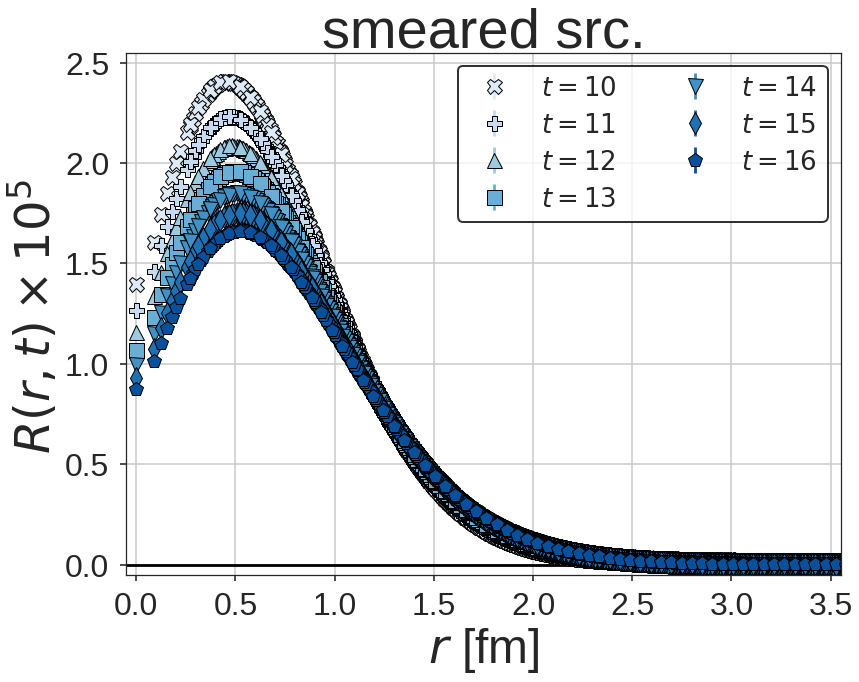}
  \caption{\label{fig:Rcorr} The $R$-correlator 
  at $t=10 - 16$    from
    the wall source (Left) and the smeared source (Right).}
\end{figure}

We first consider the behaviors of the $R$-correlator defined in  Eq.~(\ref{eq:R-correlator}).
Shown in Fig.~\ref{fig:Rcorr}
are  the $R$-correlators on the lattice with $L = 64$ at $t = 10 - 16$
from the wall source (Left) and the smeared source (Right).  
The results show strong quark-source dependence:
The $R$-correlator from the wall source ($R^\mathrm{wall}(\vec{r},t)$) is delocalized
with a weak $t$-dependence, 
while that from the smeared source ($R^\mathrm{smear}(\vec{r},t)$) is localized and has a strong $t$-dependence.
If the  $R$-correlator is saturated by the ground state,
its spatial profile should be independent of the source and its temporal profile should be simply dictated by
an overall factor, $\exp(-\Delta W_{n=0} t)$.

\begin{figure}
  \centering
  \includegraphics[width=0.49\textwidth,clip]{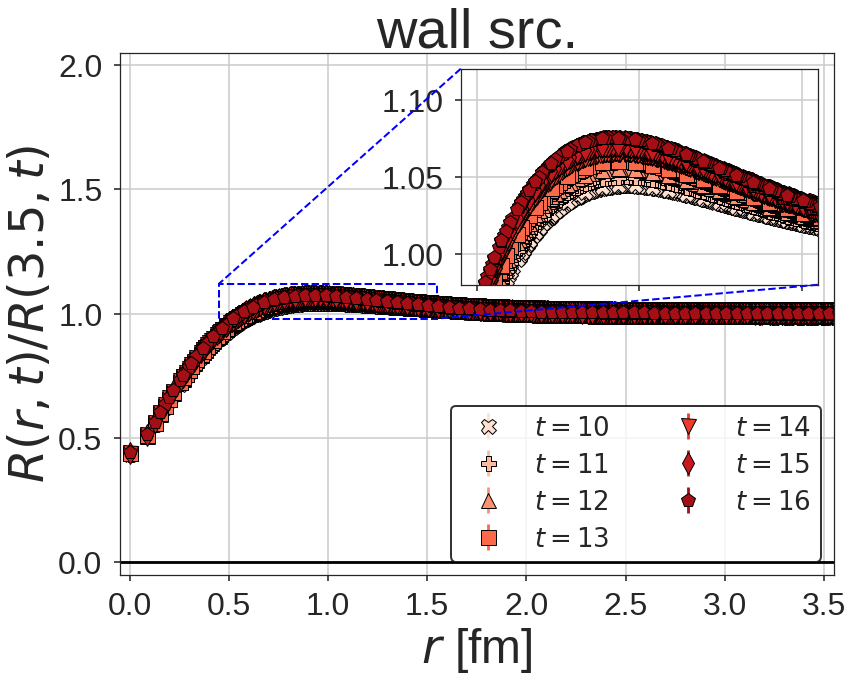}
  \includegraphics[width=0.49\textwidth,clip]{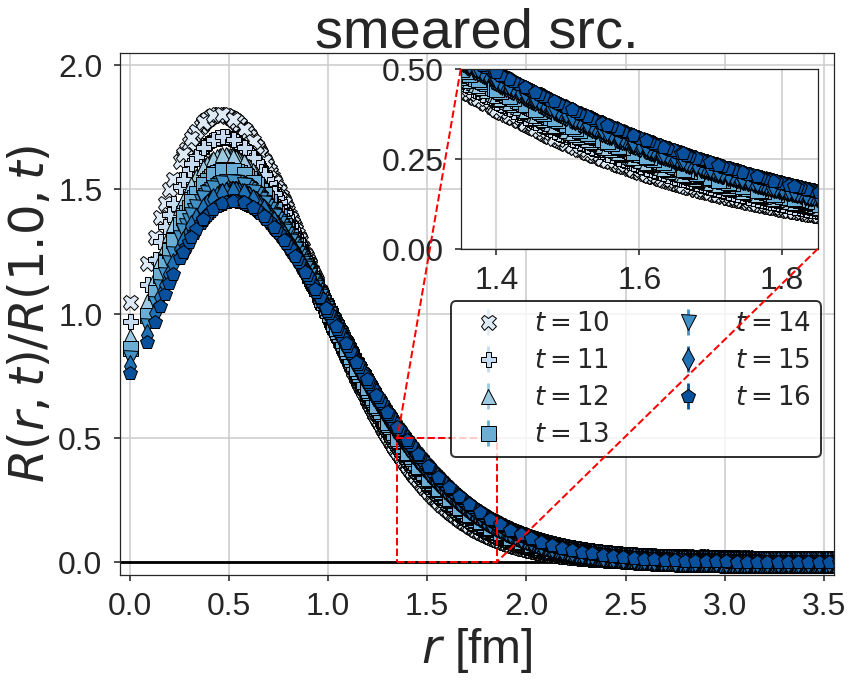}
  \caption{\label{fig:ZRcorr} The normalized $R$-correlator   at $t=10 - 16$
     from
    the wall source (Left) and the smeared source (Right).}
\end{figure}

To see more closely the $t$-dependence of the spatial profile of the $R$-correlator,
we plot $R(\vec{r},t)$ normalized to  be unity at $r=3.5$~fm for the wall source 
and at $r=1.0$~fm for the  smeared source
in Fig.~\ref{fig:ZRcorr}.
The shape of the $R$-correlator from the wall source has a weak $t$-dependence,
 which indicates that the contribution from the elastic scattering states  other than the ground state in $R^\mathrm{wall}(\vec{r},t)$ are relatively small.
  On the other hand,  the shape of $R^\mathrm{smear}(\vec{r},t)$ 
show a sizable $t$-dependence, which indicates that it has a substantial
 admixture from the several elastic scattering states.
Although the parameters $(A,B)$   of the smeared source shown in Table \ref{tab:lattice_setup} 
are tuned  to suppress the excited  states of a  single baryon,
 the same parameters are not guaranteed to 
suppress the elastic scattering states for two baryons.
Indeed,
one of the most relevant parameters   
which control the magnitudes of elastic state contributions
is the relative distance $\vec r$ between two baryons at the source, as can be illustrated from
\begin{eqnarray}
\frac{1}{L^3}\sum_{\vec x} B(\vec x) B(\vec x + \vec r) 
=  \sum_{\vec p} \tilde B(\vec p) \tilde B(-\vec p) e^{i \vec p \cdot \vec r}, \qquad
\tilde B(\vec p) \equiv \frac{1}{L^3} \sum_{\vec x} B(\vec x) e^{-i\vec p\cdot \vec x} .
\label{eq:BBoperator}
\end{eqnarray}
The smeared source operator in all previous works in the direct method
(except for~\cite{Berkowitz:2015eaa})
essentially corresponds to $\vec r=0$
and could be coupled to all elastic scattering states with almost an equal magnitude.
\footnote{
    For the studies of the meson-meson scatterings~\cite{Briceno:2017max},
    the serious systematics from the excited state contaminations
    in the plateau fitting have been widely recognized
    and the variational method~\cite{Luscher:1990ck} 
    is used with the operators analog to Eq.~(\ref{eq:BBoperator}).
}
See Ref.~\cite{Iritani:2018vfn} for more detailed studies on this point.
  
\subsection{HAL QCD potential at the leading order}

\begin{figure}
  \centering
  \includegraphics[width=0.49\textwidth,clip]{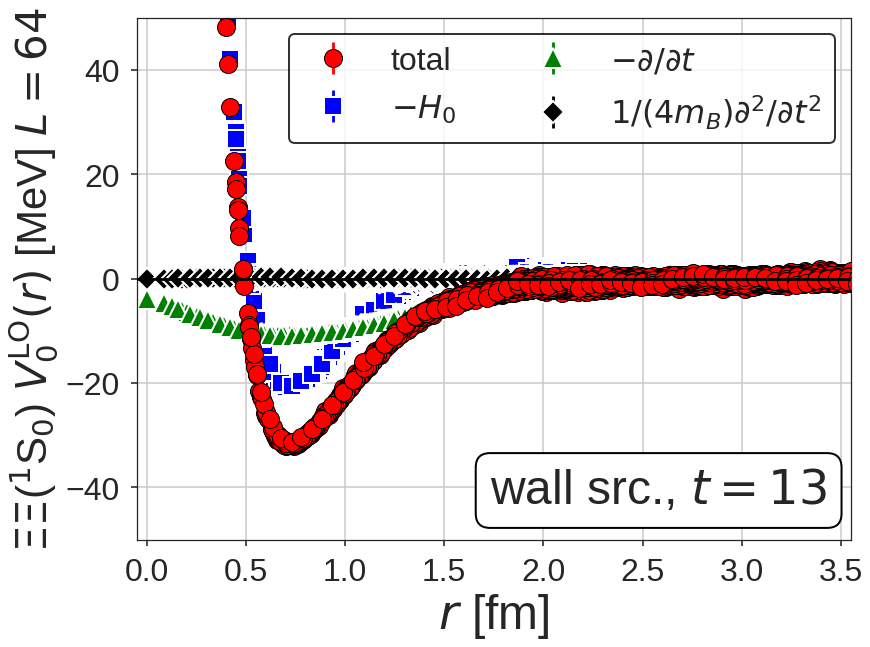}
  \includegraphics[width=0.49\textwidth,clip]{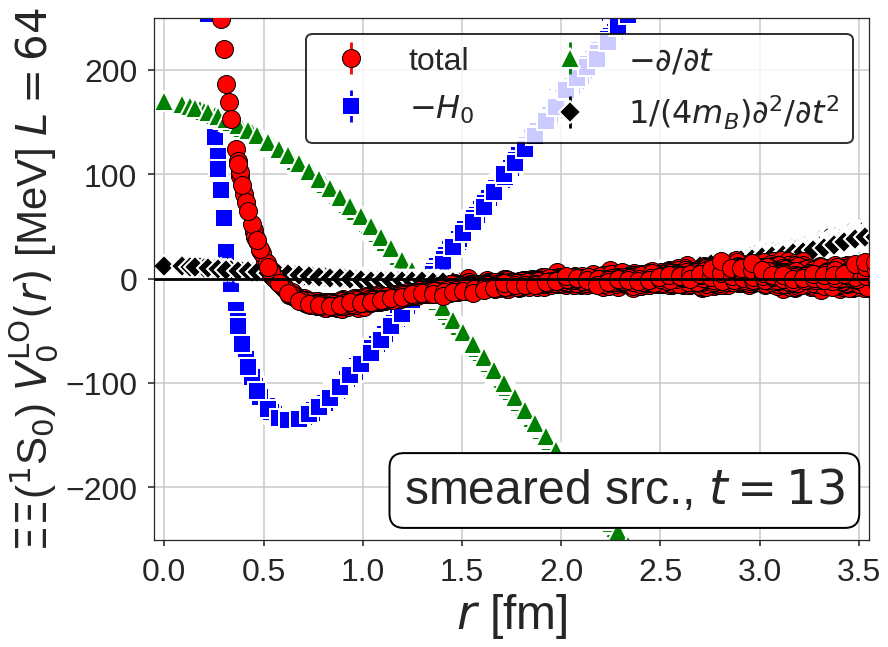}
  \caption{\label{fig:breakup}
    The potential at the leading order analysis, $V_0^\mathrm{LO}(r)$,
    (red circles)
  for the wall source (Left)
  and the smeared source (Right) at $t = 13$.
  The blue squares, green triangles and black diamonds
  denote 1st, 2nd and 3rd terms in Eq.~(\ref{eq:veff}), respectively.
}
\end{figure}

Let us now  study the potential in the HAL QCD method at the leading order, $V_0^\mathrm{LO}(r)$.
Fig.~\ref{fig:breakup} shows the one for   $\Xi\Xi$($^1$S$_0$)
and its breakups ($H_0$, $\partial/\partial t$ and $\partial^2/\partial t^2$ terms in Eq.~(\ref{eq:veff}))
on $L = 64$ at $t = 13$  from the wall source (Left) and the smeared source (Right).
For the wall source,
the $H_0$ term is dominant with sizable contributions from the $\partial/\partial t$ term,
  while the $\partial^2/\partial t^2$ term is negligible. 
  The $\partial/\partial t$ term is not constant as a function of $r$, which  
  indicates that there exist small but non-negligible contributions from the excited states 
  in $R^\mathrm{wall}(\vec{r},t)$.
For the smeared source, on the other hand,
all terms are important.
In particular, the $\partial/\partial t$ term (green triangles) 
shows substantial $r$-dependence indicating 
 large contributions from the excited states  in the smeared source.
However, such dependence is cancelled by the $H_0$ term (blue squares) and is 
 further corrected by the  $\partial^2/\partial t^2$ term (black diamonds).
 The final results (red circles) with the smeared source and the wall source 
 show qualitatively similar behaviors, i.e., 
the repulsive core at the short distance and 
the attractive pocket at the intermediate distance. 
This illustrates  that  the time-dependent HAL QCD method works 
well for extracting the $\Xi\Xi$ potential  irrespective of the source structures.

\begin{figure}
  \centering
  \includegraphics[width=0.49\textwidth,clip]{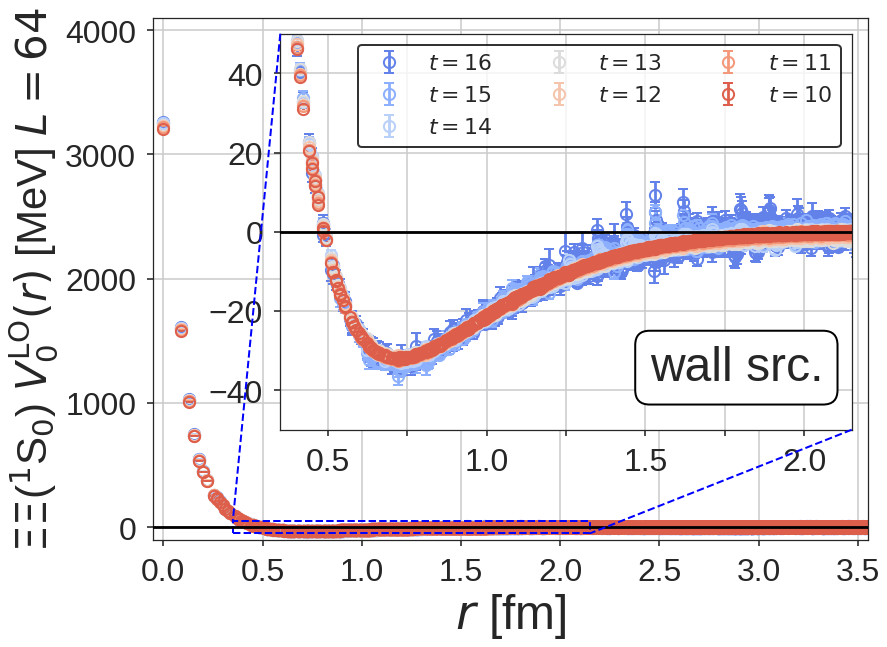}
  \includegraphics[width=0.49\textwidth,clip]{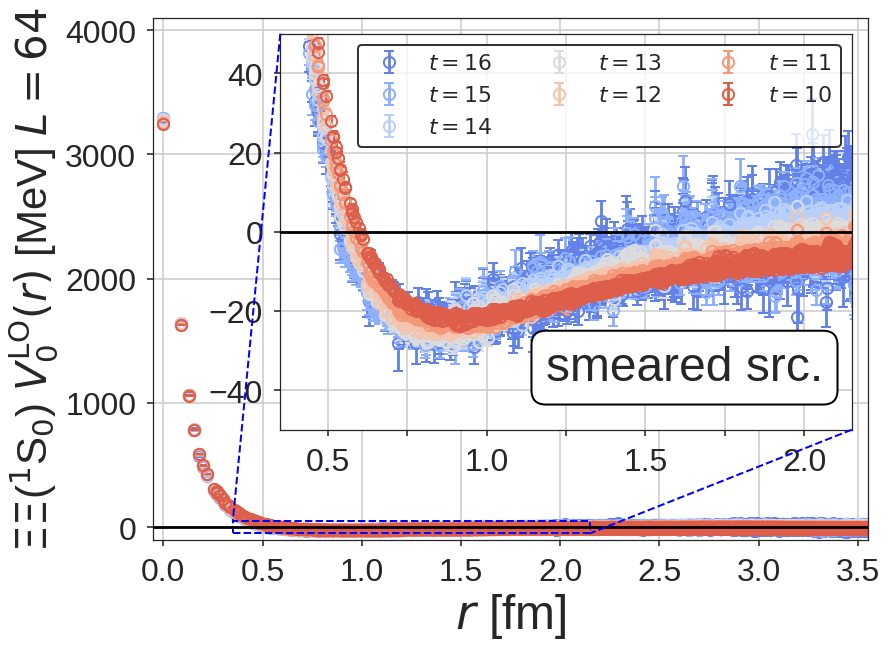}
  \caption{
    The potential at  the leading order analysis, $V_0^\mathrm{LO}(r)$,
    for the wall source (Left) and the smeared source (Right)
    at $t = 10 - 16$.
    \label{fig:wall_vs_smear}
}
\end{figure}

\begin{figure}
  \centering
  \includegraphics[width=0.49\textwidth,clip]{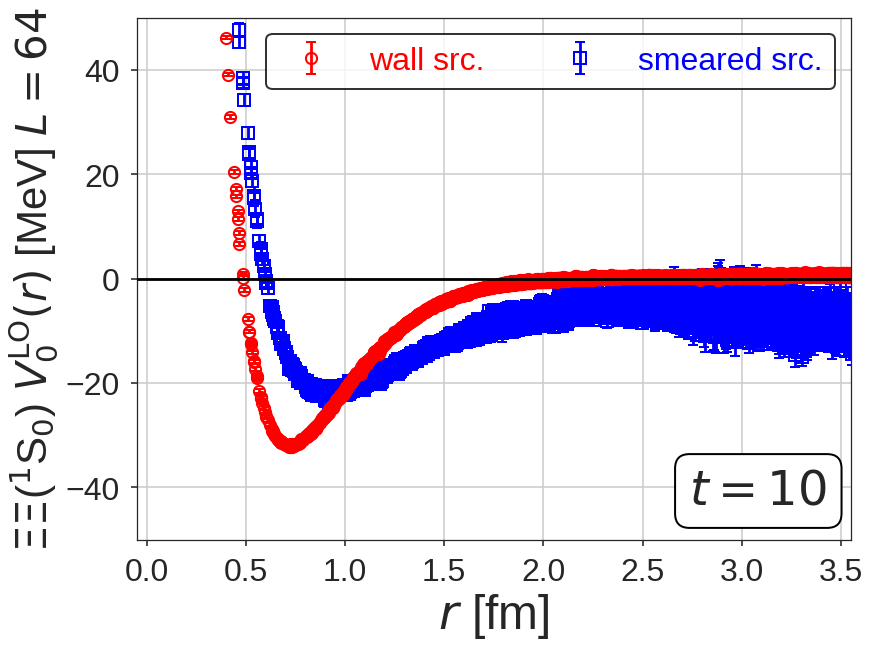}
  \includegraphics[width=0.49\textwidth,clip]{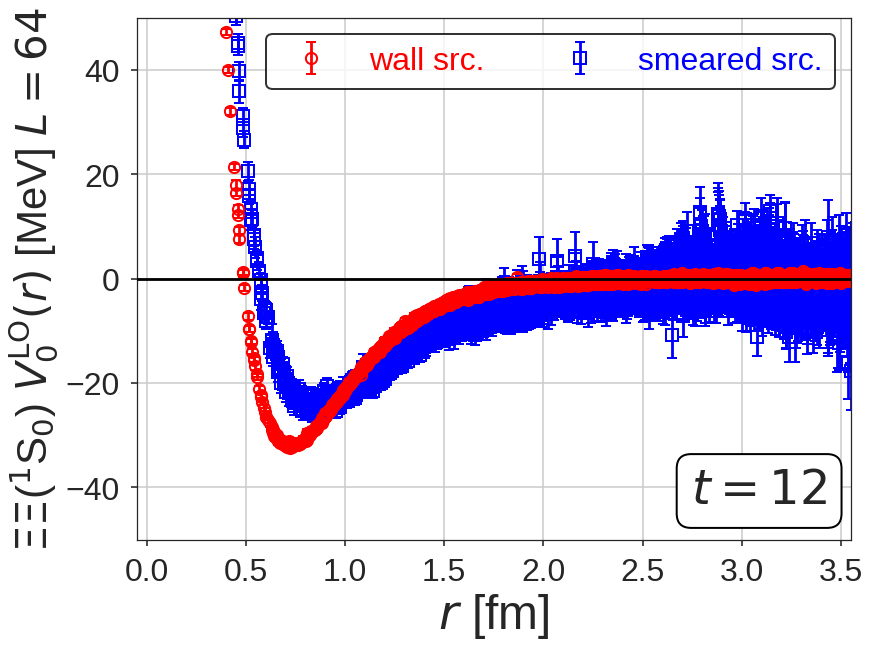}
  \includegraphics[width=0.49\textwidth,clip]{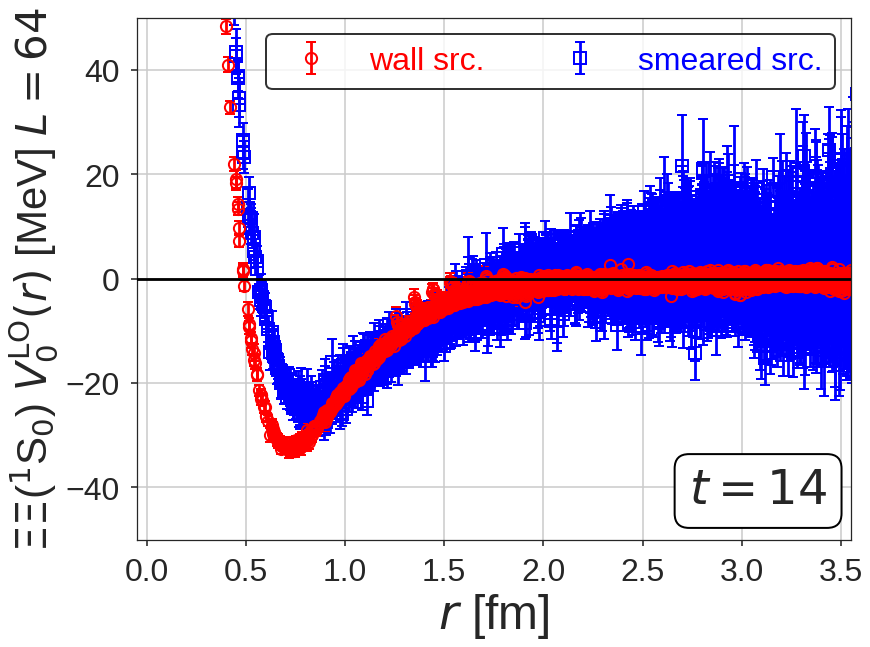}
  \includegraphics[width=0.49\textwidth,clip]{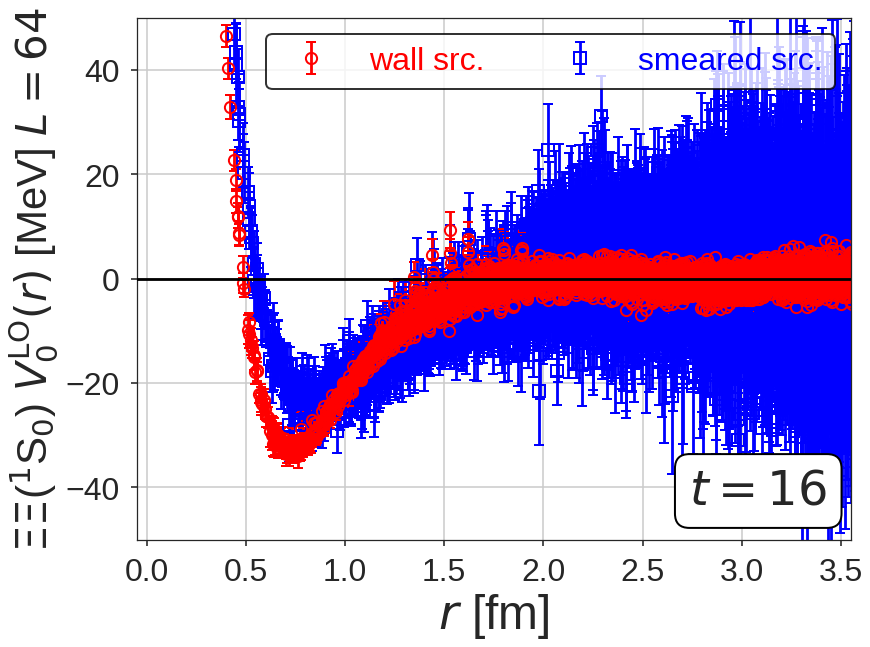}
  \caption{
    A comparison of the potential at the leading order analysis, $V_0^\mathrm{LO}(r)$,
      between the wall source (red circles) and the smeared source (blue squares)
      at $t = 10, 12, 14, 16$.
    \label{fig:wall_vs_smear_comp}
}
\end{figure}

Shown in Fig.~\ref{fig:wall_vs_smear} is a 
comparison among the LO potentials  ($V_0^\mathrm{LO}(r)$)
 for  different  $t$ in each source.
For the wall source, the potentials at $t=10-16$ are consistent with each other
within statistical errors,
while those from  the smeared source show the detectable $t$-dependence.
 Shown in Fig.~\ref{fig:wall_vs_smear_comp} is a comparison of $V_0^\mathrm{LO}(r)$
between two sources at  $t = 10, 12, 14, 16$.   As $t$ increases,
the LO potential from the smeared source gradually converges to that from the wall source.
The relatively large $t$-dependence of the potentials from the smeared source
as well as the remaining small discrepancy of potentials  between two sources
even at $t = 16$
indicate that
the N$^2$LO analysis in the derivative expansion is necessary
to understand the data from the smeared source.
This is a natural consequence of the fact that
the N$^2$LO contributions in $V_0^\mathrm{LO}(r)$,
$\nabla^2 R(\vec{r},t)/R(\vec{r},t)$ ($\propto H_0$ term) in Eq.~(\ref{eq:vnlo}),
is much more significant in the smeared source than the wall source as shown in Fig.~\ref{fig:breakup}.

\subsection{HAL QCD potential at the next-to-next-to-leading order}

We next apply the N$^2$LO analysis in the derivative expansion to $R$-correlators for both sources.
The potential at the LO analysis, $V_0^\mathrm{LO}(r)$,
and those at the N$^2$LO analysis, $V_0^\mathrm{N^2LO}(r)$, $V_2^\mathrm{N^2LO}(r)$,
satisfy  the linear equations given by
\begin{eqnarray}
  \{ V_0^\mathrm{N^2LO}(r)  + V_2^\mathrm{N^2LO}(r)\nabla^2 \} R^\mathrm{source}(\vec{r},t) =  V_0^\mathrm{LO(source)}(r) R^\mathrm{source}(\vec{r},t) ,
  \label{eq:nlo_pot}
\end{eqnarray}
where source = wall or smear.

To extract $V_{0,2}^\mathrm{N^2LO}(r)$, we first consider the 
following relation derived from  Eq.~(\ref{eq:nlo_pot}),
\begin{eqnarray}
D \times V_2^\mathrm{N^2LO}(r)  = V_0^\mathrm{LO(wall)}(r) - V_0^\mathrm{LO(smear)}(r) ,
   \label{eq:nlo_pot2}
\end{eqnarray}
with $D \equiv 
\nabla^2 R^\mathrm{wall}(\vec{r},t)/R^\mathrm{wall}(\vec{r},t)
-\nabla^2 R^\mathrm{smear}(\vec{r},t)/R^\mathrm{smear}(\vec{r},t)$.
In order to avoid numerical instabilities caused by nearly zeros of $D$ when we 
divide the right hand side of Eq.~(\ref{eq:nlo_pot2}), 
we extract $V_2^\mathrm{N^2LO}(r)$ directly from Eq.~(\ref{eq:nlo_pot2})
with 
a fitting function,
$V_2^\mathrm{N^2LO}(r)  = b_{1} e^{-b_{2}(r-b_{{3}})^2} + b_{{4}} e^{-b_{{5}}(r-b_{{6}})^2}$
at each $t$.
Once $V_2^\mathrm{N^2LO}(r)$ is obtained, 
$V_0^\mathrm{N^2LO}(r)$ can be determined from Eq.~(\ref{eq:vnlo}).

\begin{figure}
  \centering
  \includegraphics[width=0.49\textwidth,clip]{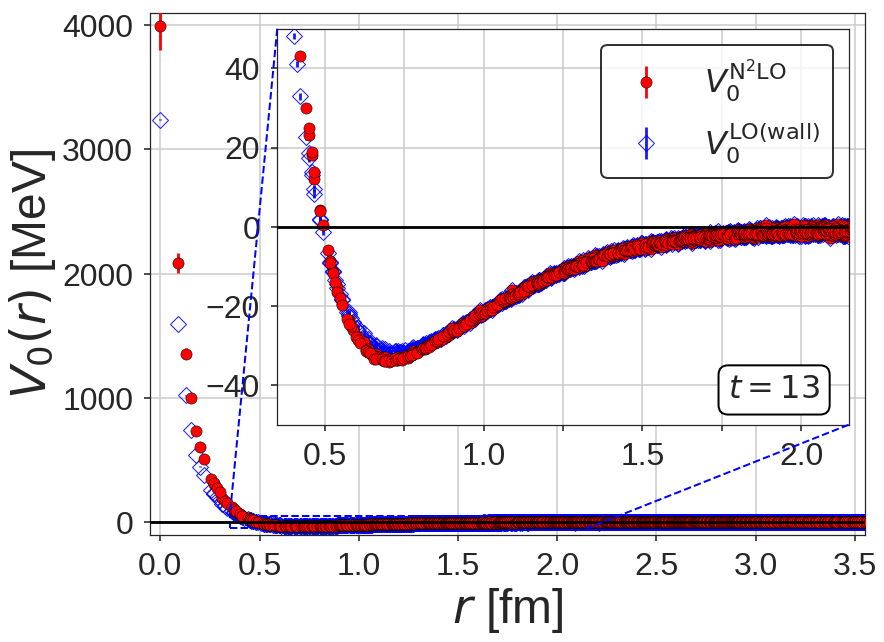}
  \includegraphics[width=0.49\textwidth,clip]{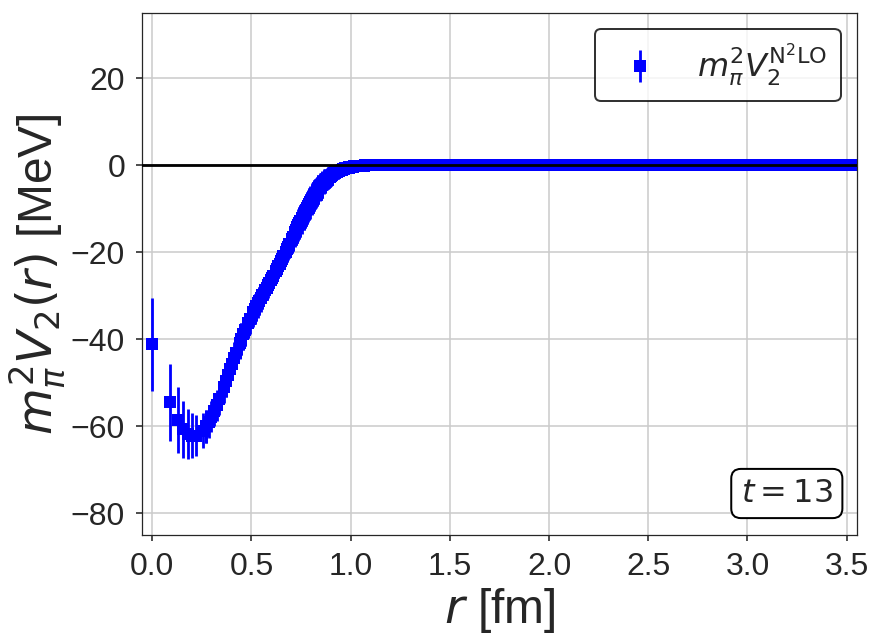}
  \caption{
    \label{fig:vlo_vnlo}
    (Left) 
      The LO potential at the N$^2$LO analysis, $V_0^\mathrm{N^2LO}(r)$ (red circles),
      together with the potential at the LO analysis for the wall source, $V_0^\mathrm{LO(wall)}(r)$
      (blue diamonds) at $t=13$.
    (Right) The N$^2$LO potential at the N$^2$LO analysis, $V_2^\mathrm{N^2LO}(r)$,
    multiplied by $m_\pi^2$.
  }
\end{figure}

Fig.~\ref{fig:vlo_vnlo} shows the $V_0^\mathrm{N^2LO}(r)$ 
together with the $V_0^\mathrm{LO(wall)}(r)$ (Left), 
and the $V_2^\mathrm{N^2LO}(r) $ (Right) on $L = 64$ at $t= 13$.
 We multiply $V_2^\mathrm{N^2LO}(r)$ by $m_\pi^2$ to make its mass dimension $+1$
for a comparison to $V_0(r)$'s.
 We find that $V_0^\mathrm{N^2LO}(r)$ agrees well  with the 
$V_0^\mathrm{LO(wall)}(r) $ except at short distances.
We also find that  $V_2^\mathrm{N^2LO}(r) $ is localized within the range of 1~fm, which is much shorter than
the range of  $V_{0}^{{\mathrm{LO(wall),N^2LO}}}(r)$.
We note here that the negative sign of $V_2^\mathrm{N^2LO}(r)$ does not necessarily 
 imply attraction,  since the N$^2$LO potential is  given by $V_2^\mathrm{N^2LO}(r)\nabla^2$.

As already mentioned,
$\nabla^2 R(\vec{r},t)/R(\vec{r},t)$ from the smeared source is much larger than that of the wall source
(see Fig.~\ref{fig:breakup}).
Intuitively, this is because $R^\mathrm{smear}(r,t)$ ($R^\mathrm{wall}(r,t)$)
  contains larger (smaller) contributions from excited states
  and thus is more (less) sensitive to higher order terms in the derivative expansion of the potential.
Therefore, the N$^2$LO analysis is mandatory  for the smeared source,
while the LO analysis for the wall source leads to the
potential which is almost identical to  $V_0^\mathrm{N^2LO}(r)$.

\begin{figure}
  \centering
  \includegraphics[width=0.49\textwidth,clip]{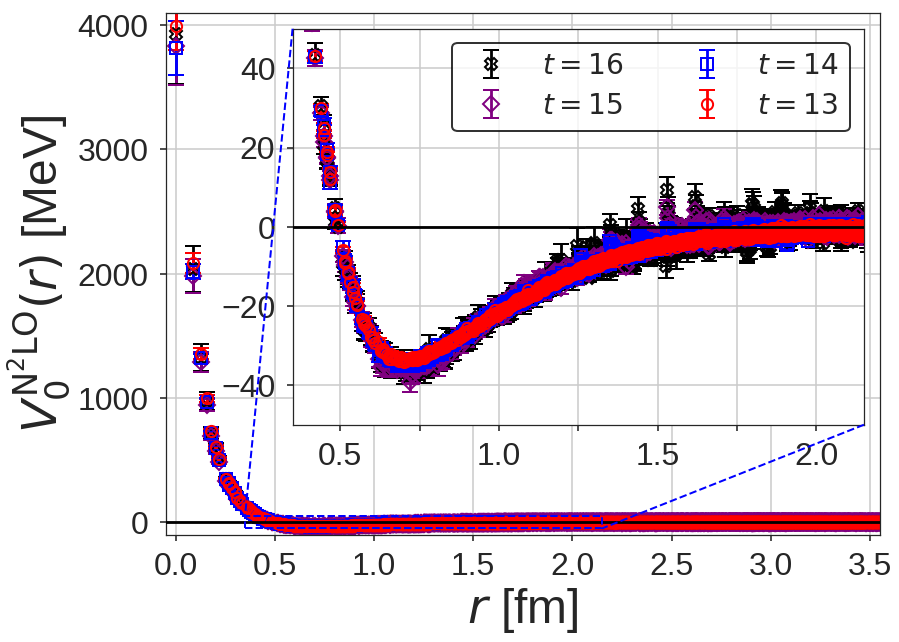}
  \includegraphics[width=0.49\textwidth,clip]{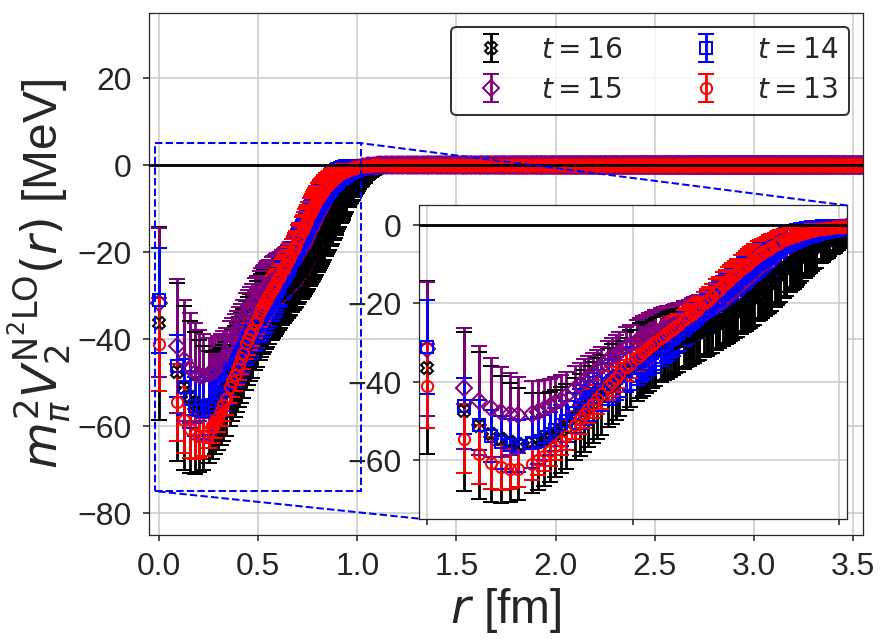}
  \caption{
    \label{fig:vlo_vnlo_t_dep}
    The LO (Left) and N$^2$LO (Right) potentials
    at the N$^2$LO analysis in the range of $t = 13 - 16$.
  }
\end{figure}

Shown in Fig.~\ref{fig:vlo_vnlo_t_dep} is
the $t$-dependence of $V_{0,2}^\mathrm{N^2LO}(r) $ in the range of $t = 13 - 16$.
Since appreciable $t$-dependence is not seen within the error bars,
the N$^4$LO contribution is expected to be small.

\subsection{Effect of the Inelastic states}

\begin{figure}
  \centering
  \includegraphics[width=0.49\textwidth,clip]{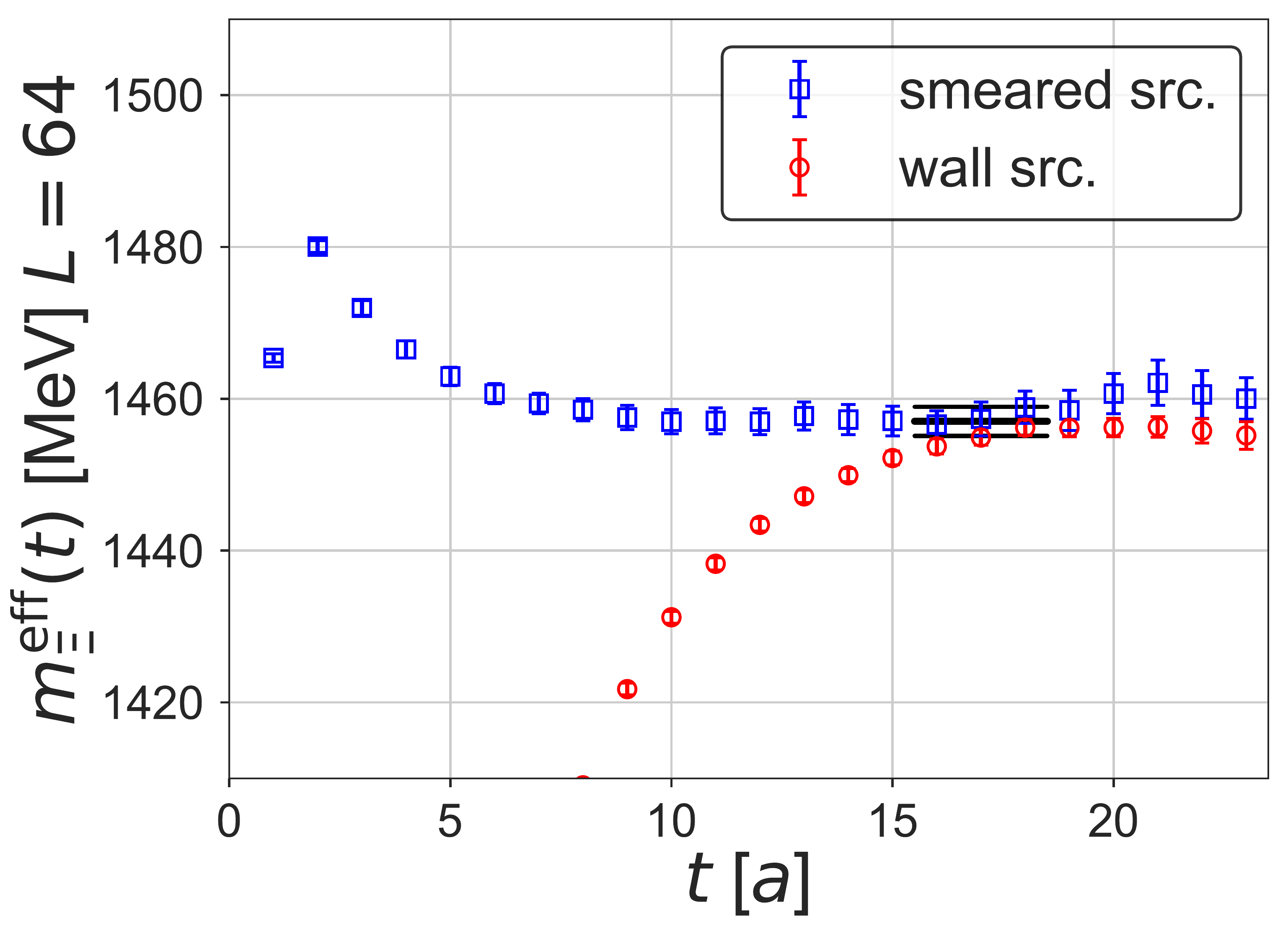}
  \includegraphics[width=0.49\textwidth,clip]{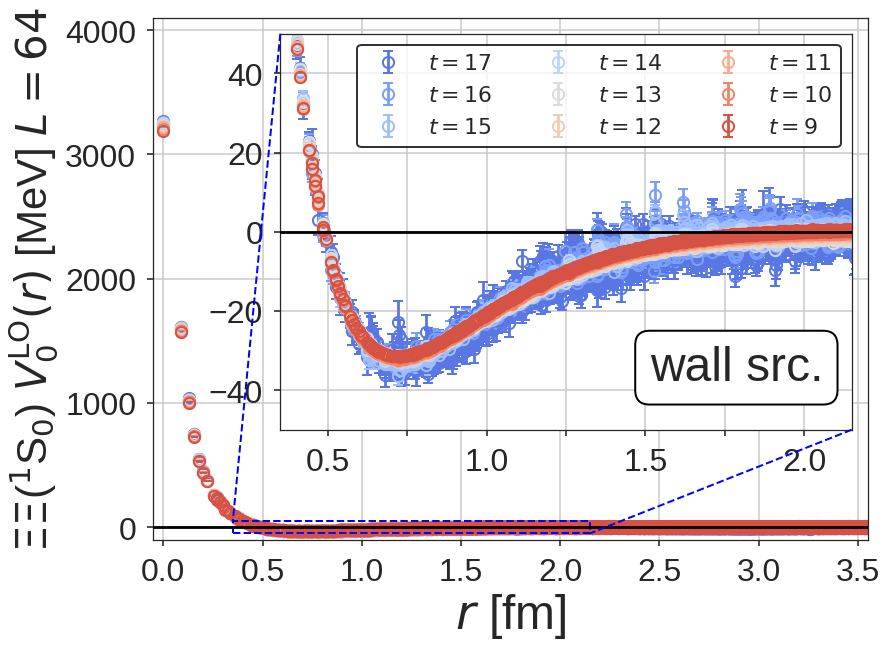}
  \caption{
    \label{fig:wall_plateau}
    (Left) The effective mass of a single baryon $\Xi$
    for the wall source (red circles) and the smeared source (blue squares).
    (Right) The potential at the LO analysis, $V_0^\mathrm{LO}(r)$, for the wall source at $t=9 - 17$.
  }
\end{figure}

Fig.~\ref{fig:wall_plateau}~(Left) compares the effective mass of  a single $\Xi$ for two sources.
 The smeared source is tuned to have a large overlap with the ground state of a single baryon,
so that the corresponding effective mass shows a plateau at an earlier time
 than the case of the wall source.  Eventually, the plateaux for the single $\Xi$ from two different 
 sources converge at $t \gtrsim 16$.
Shown in Fig.~\ref{fig:wall_plateau}~(Right) is the $\Xi \Xi$
 potential at the LO analysis for the wall source in the range of $t = 9 -17$.
 Unlike the case of the single $\Xi$, the resultant potential is stable for $t$ much less than $16$, 
 suggesting that the systematic error originating from the inelastic contributions of the  single-baryon
  cancels largely between the numerator and the denominator of the $R$-correlator  for the wall source. 

\subsection{Effect of the finite volume}

\begin{figure}
  \centering
  \includegraphics[width=0.49\textwidth,clip]{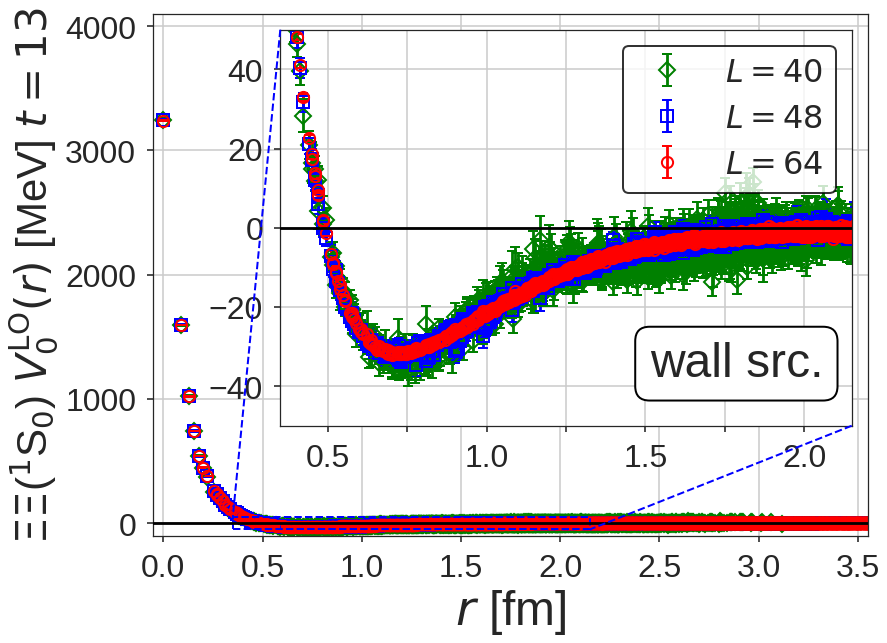}
  \caption{
    \label{fig:veff_vol_dep}
    The potential at the LO analysis, $V_0^\mathrm{LO}(r)$, for the wall source on $L = 40, 48$ and $64$
    at $t = 13$.
  }
\end{figure}

In Fig.~\ref{fig:veff_vol_dep}, we  show the volume dependence of 
the potential at  the LO analysis for the wall source  at $t = 13$ with $L = 40$, $48$ and $64$.
All the  potentials  are consistent with each other within statistical errors. This
indicates that the artifact due to finite volume  is negligible for the potential, mainly because the
potential is short ranged.

\section{Scattering phase shifts}

In the previous section, we examine systematic uncertainties
on the HAL QCD potential.
In this section, we examine how these systematic uncertainties 
 affect the physical observables such as  
the scattering phase shifts, in particular the effect of 
the derivative expansion. To calculate the scattering phase shifts, $\delta_0(k)$,
we first fit the potentials 
by a sum of  Gaussians,
$  V_0^\mathrm{LO(wall), N^2LO}(r) = \sum_{n=1,3,5,7}  a_n e^{-a_{n+1} r^2}$  and 
$  V_{2}^\mathrm{N^2LO}(r)= \sum_{n=1,4}  b_n e^{-b_{n+1}(r-b_{n+2} )^2} $.
Resulting  parameters are summarized in Table~\ref{tab:fit_params}.

\begin{table}[h]
  \centering
  \begin{tabular}{|c|c|c||c|c|}
    \hline 
    \hline
    & $V_0^\mathrm{LO(wall)}(r)$ & $V_0^\mathrm{N^2LO}(r)$ & & $V_2^\mathrm{N^2LO}(r)$ \\
    \hline
    $a_1$ & $0.8759 \pm 0.0270$ & $1.1426 \pm 0.0621$& $b_1 $  & $-0.5291 \pm 0.0418$ \\
    $a_2$ & $1.2040 \pm 0.0317$ & $0.9332 \pm 0.0871$& $b_2$ & $0.0757 \pm 0.0162$ \\
    $a_3$ & $0.4261 \pm 0.0128$ & $0.4245 \pm 0.0397$& $b_3   $  & $2.195 \pm 0.333$ \\
    $a_4$ & $0.3028 \pm 0.0217$ & $0.2358 \pm 0.0382$& $b_4$ & $-0.1091 \pm 0.0194$ \\
    $a_5$ & $0.2010 \pm 0.0124$ & $0.2415 \pm 0.0410$& $b_5$         & $0.2177 \pm 0.0633$ \\
    $a_6$ & $0.07373 \pm 0.00364$ & $0.07876 \pm 0.00646$& $b_6$     & $7.025 \pm 0.464$ \\
    $a_7$ & $-0.02922 \pm 0.00148$ & $-0.03005 \pm 0.00159$ & & \\
    $a_8$ & $0.008977 \pm 0.000456$ & $0.009107 \pm 0.000467$ & & \\
    \hline
    \hline
  \end{tabular}
  \caption{Summary of fitting parameters for the LO and N$^2$LO potentials in 
    the lattice unit at $t = 13$.
    The fitting range is $r \in [0, 3.5]$ fm, and 
    $\chi^2/\mathrm{dof}$ are  
    1.14, 1.01 and 0.0019
    for $V_0^\mathrm{LO(wall)}(r)$, $V_0^\mathrm{N^2LO}(r)$ and $V_2^\mathrm{N^2LO}(r)$, respectively. 
  }
  \label{tab:fit_params}
\end{table}

\begin{figure}
  \centering
  \includegraphics[width=0.49\textwidth,clip]{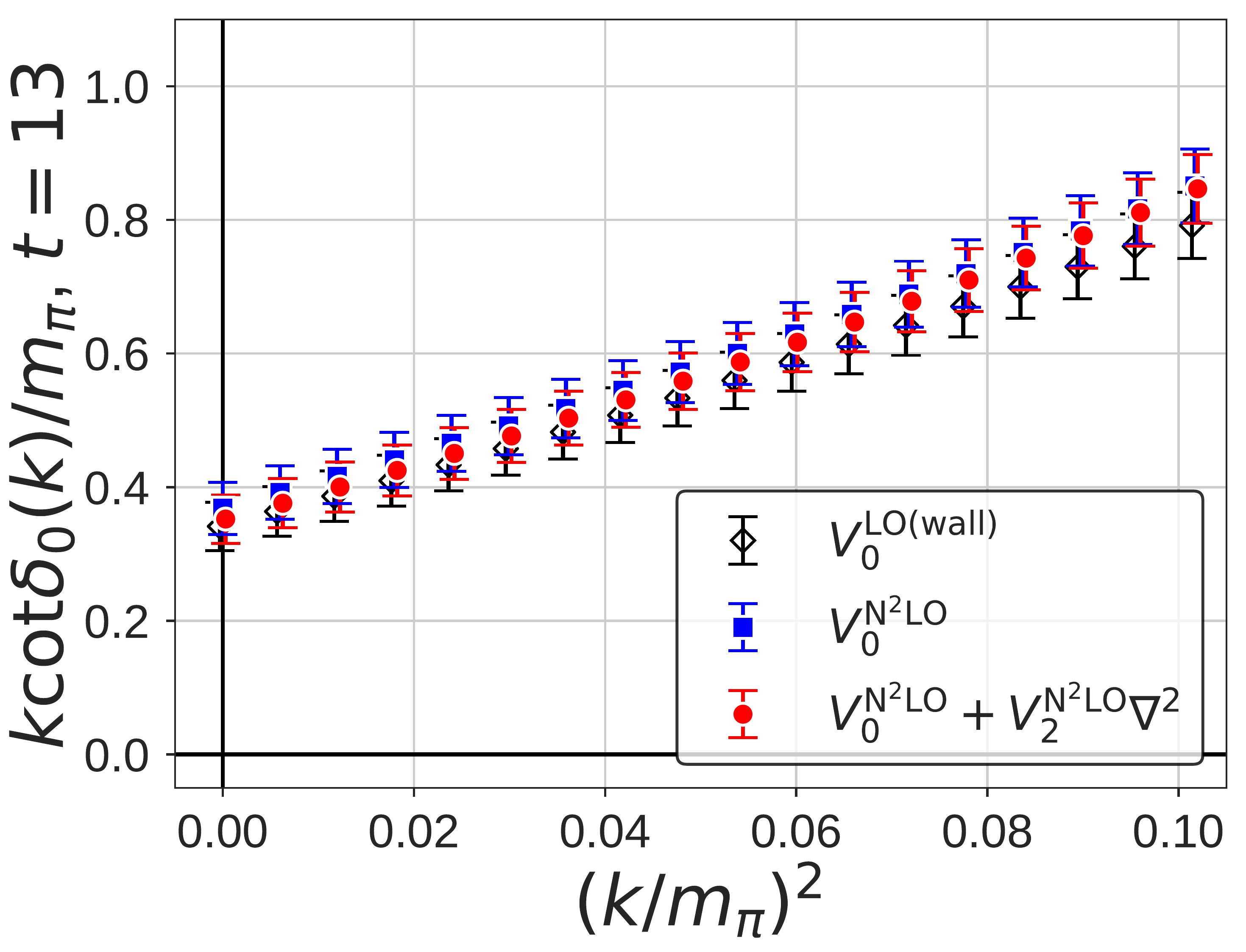}
  \includegraphics[width=0.49\textwidth,clip]{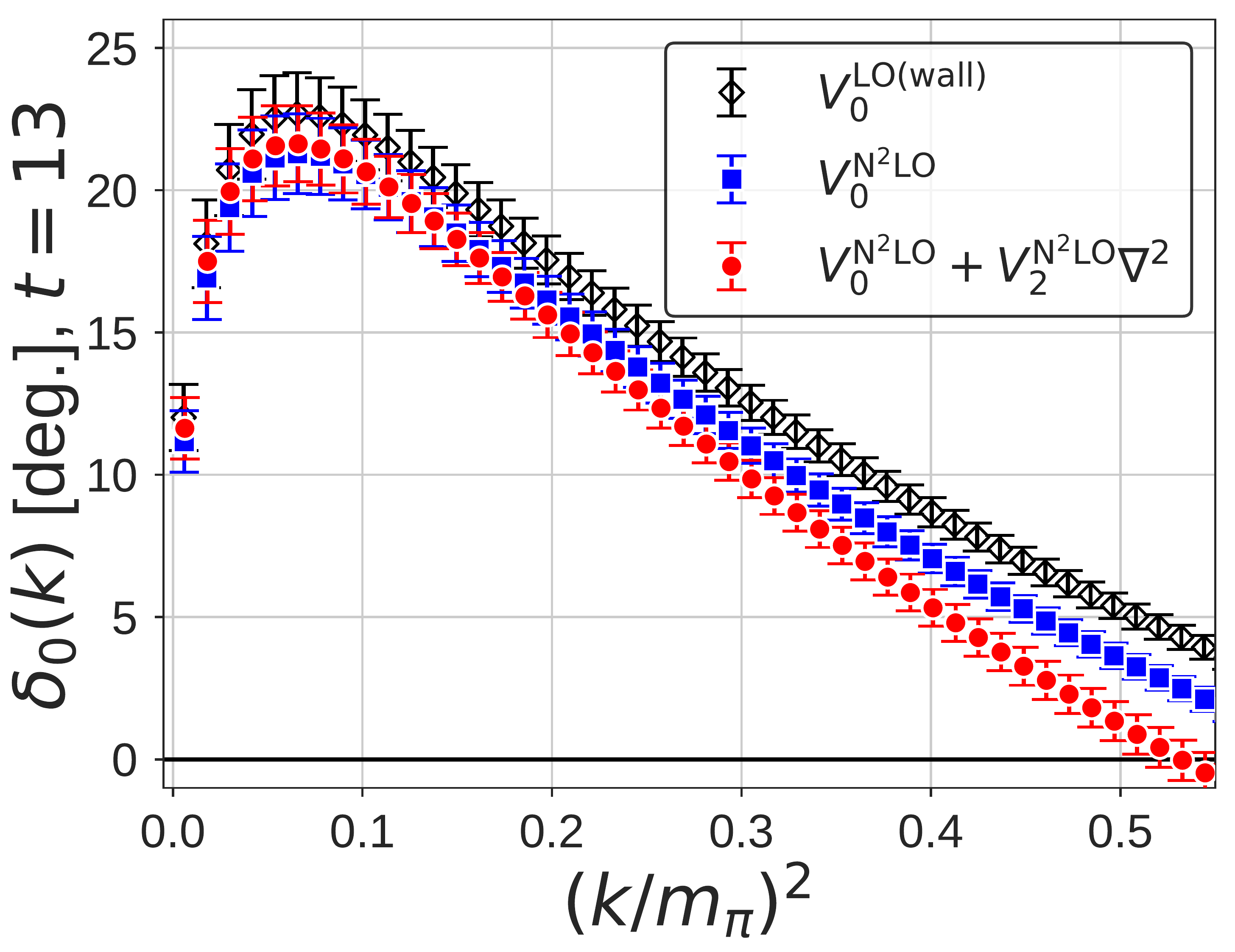}
  \caption{
  \label{fig:scat_phase}
  The scattering phase shifts in the form of $k\cot\delta_0(k)/m_\pi$ (Left) and $\delta_0(k)$ (Right)
  from
  $V_0^\mathrm{LO(wall)}(r)$ (black diamonds), 
  $V_0^\mathrm{N^2LO}(r)$ (blue squares) and $V_0^\mathrm{N^2LO}(r) + V_2^\mathrm{N^2LO}(r)\nabla^2$ (red circles)  at $t = 13$.
}
\end{figure}

In Fig.~\ref{fig:scat_phase}, we show the comparison
of the scattering phase shifts from $V_0^\mathrm{LO(wall)}(r)$,
$V_0^\mathrm{N^2LO}(r)$ and $V_0^\mathrm{N^2LO}(r) + V_2^\mathrm{N^2LO}(r)\nabla^2$
at $t=13$.
At low energies (Fig.~\ref{fig:scat_phase}~(Left)), the N$^2$LO correction is found to be negligible, 
showing not only that the derivative expansion converges well 
but also that the LO analysis for the wall source is sufficiently good at low energies.
The N$^2$LO correction becomes non-negligible only at high energies as shown in Fig.~\ref{fig:scat_phase}~(Right)
\footnote{We discuss the magnitude of the N$^2$LO correction in the potential
at high energies in Appendix~\ref{app:N2LOcorr}.}.
 We note that $(k/m_\pi)^2 = 0.5$ corresponds to  the energy from the threshold as
  $\Delta E \equiv W - 2m_B \simeq 90$~MeV.
The good convergence of the derivative expansion
has been also observed for the $NN$ systems in the $^1$S$_0$ and $^3$S$_1$ channels
in quenched QCD with $m_{\pi} \simeq 530$ MeV~\cite{Murano:2011nz} 
and the $I=2$ $\pi\pi$ system in (2+1)-flavor QCD with $m_{\pi} \simeq 870$ MeV~\cite{Kawai:2017goq}.

\begin{figure}
  \centering
  \includegraphics[width=0.49\textwidth,clip]{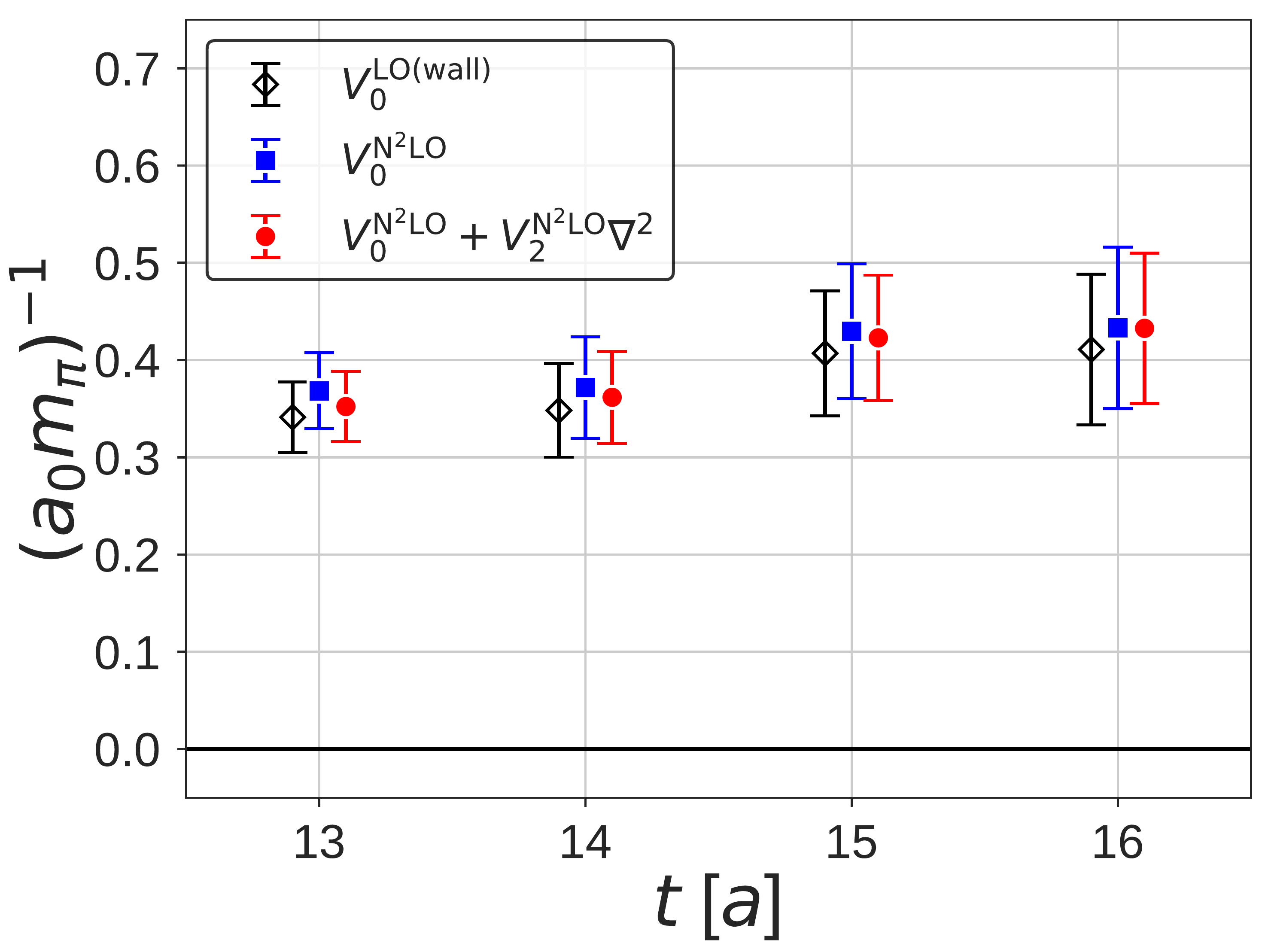}
  \caption{
  \label{fig:scat_length}
    The scattering length $a_0$ in the form of $(a_0 m_\pi)^{-1}$ 
    from $V_0^\mathrm{LO(wall)}(r)$ (black diamonds) ,
    $V_0^\mathrm{N^2LO}(r)$ (blue squares) and
    $V_0^\mathrm{N^2LO}(r) + V_2^\mathrm{N^2LO}(r)\nabla^2$ (red circles)
    at $t = 13-16$.
  }
\end{figure}

The scattering length $a_0$ obtained through
$\lim_{k\rightarrow 0} k\cot\delta_0(k) = 1/a_0$
from $V_0^\mathrm{LO(wall)}(r)$, $V_0^\mathrm{N^2LO}(r)$ and $V_0^\mathrm{N^2LO}(r) + V_2^\mathrm{N^2LO}(r)\nabla^2$
at $t = 13-16$ is shown in Fig.~\ref{fig:scat_length}.
The result indicates that the scattering length is almost insensitive to the degrees of the approximation
but has a small variation in $t$, which is, however, within statistical errors.  
We thus conclude that
the systematic errors from the derivative expansion and the
inelastic state contaminations are well under control for this observable.
Numerical values for the scattering length are summarized in Table~\ref{tab:scat_length},
where the central value and statistical errors are evaluated at $t=13$ and 
the systematic errors are estimated from the $t$-dependence among $t = 13-16$.
We have checked that alternative fitting functions of the potential such as the 
combination of two Gaussians + (Yukawa)$^2$ form 
as employed in ~\cite{Aoki:2012tk, Yamada:2015cra} 
give results consistent with those from the present fitting function within errors. 

\begin{table}
  \centering
  \begin{tabular}{|c|c|c|c|}
    \hline
    \hline
    & $V_0^\mathrm{LO(wall)}(r)$ & $V_0^\mathrm{N^2LO}(r)$ & $V_0^\mathrm{N^2LO}(r) + V_2^\mathrm{N^2LO}(r)\nabla^2$ \\
    \hline
    $(a_0 m_\pi)^{-1}$ &
    0.341(36){(${}^{+70}_{-0}$)} &
    0.368(39){(${}^{+65}_{-0}$)} & 
    0.352(36){(${}^{+80}_{-0}$)} \\
    \hline
    \hline
  \end{tabular}
  \caption{The scattering length $a_0$ in the form of $(a_0 m_\pi)^{-1}$ from 
    $V_0^\mathrm{LO(wall)}(r)$, $V_0^\mathrm{N^2LO}(r)$ and 
    $V_0^\mathrm{N^2LO}(r) + V_2^\mathrm{N^2LO}(r)\nabla^2$.  
    The central values and statistical errors (in the first parenthesis) are evaluated at $t = 13$, 
    while the systematic errors  (in the second) are estimated using the potentials at $t = 14, 15, 16$.
}
  \label{tab:scat_length}
\end{table}

\section{Finite volume formula and effective range expansion}

Before closing the paper, we discuss the relation among
the energy spectrum,
the L\"uscher's finite volume formula and the effective range expansion (ERE).
Once the energy shift of the two-body system on a finite volume is measured,
the scattering phase shift is obtained by the L\"uscher's  formula as
\begin{equation}
  k\cot\delta_0 (k) = \frac{1}{\pi L} \sum_{\vec{n} \in \mathbf{Z}^3} \frac{1}{|\vec{n}|^2 - (kL/2\pi)^2},
\end{equation}
where  $k^2$ is related to the energy shift on a finite volume as
  $\Delta E_L = 2\sqrt{m_B^2 + k^2} - 2m_B$.
For the attractive  interaction, $k^2 $ can be negative on a finite volume.
  Note that the poles of the $S$-matrix with $k\cot\delta_0(k) = - \sqrt{-k^2}$ in the infinite volume
  correspond to the bound states.
For the unbound two-body system, the asymptotic behavior of $\Delta E_L$ 
 for large $L$ reads
\begin{equation}
  \Delta E_L \simeq - \frac{2\pi a_0}{\mu L^3} 
  \left[ 1 + c_1 \frac{a_0}{L} + c_2 \left( \frac{a_0}{L} \right)^2 \right] + \mathcal{O}(L^{-6}) ,
  \label{eq:L_dep_scat}
\end{equation}
with the reduced mass $\mu$, the scattering length $a_0$,
$c_1 = -2.837297$, and $c_2 = 6.375183$~\cite{Luscher:1985dn, Luscher:1990ux}.

\begin{figure}
  \centering
  \includegraphics[width=0.49\textwidth,clip]{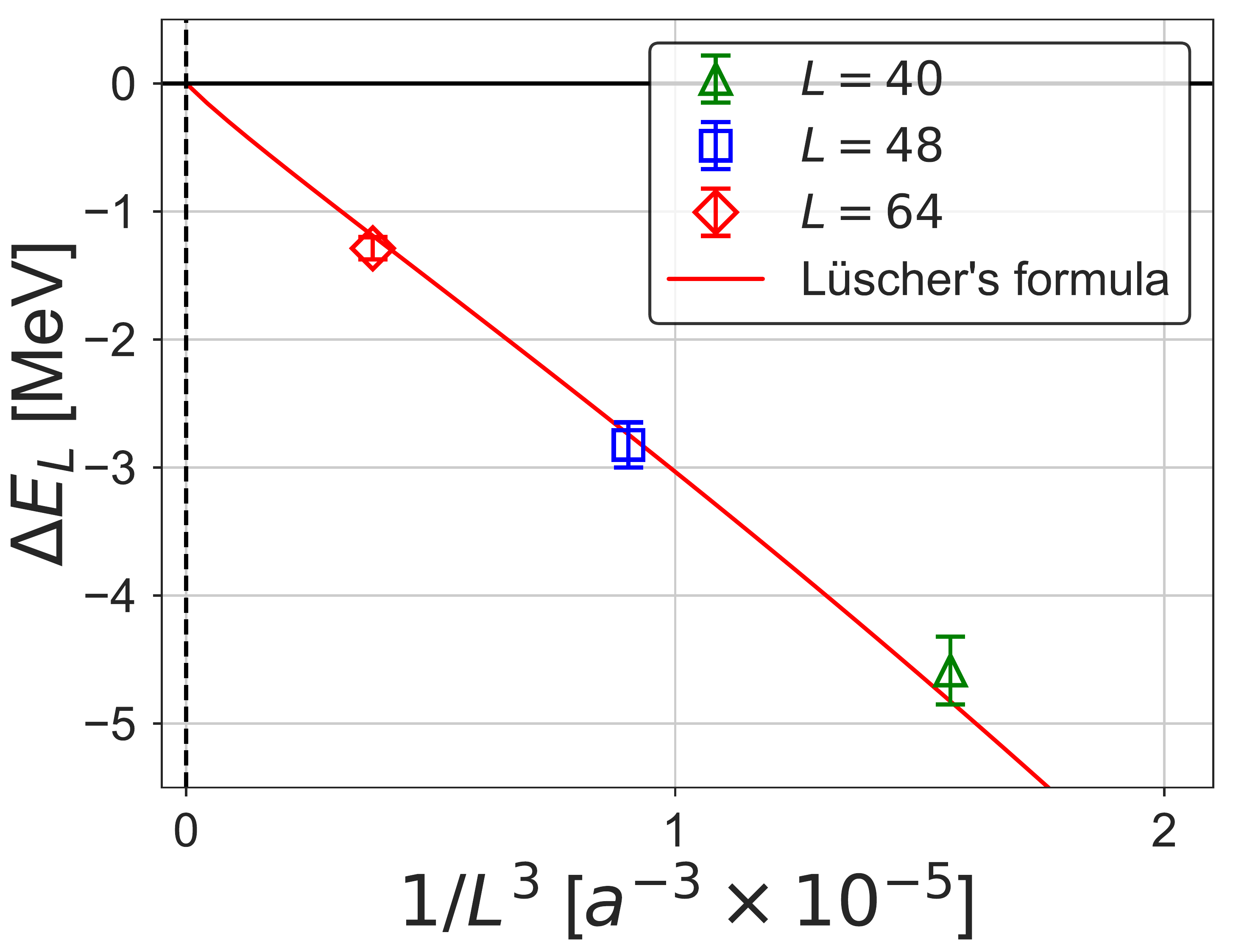}
  \includegraphics[width=0.50\textwidth,clip]{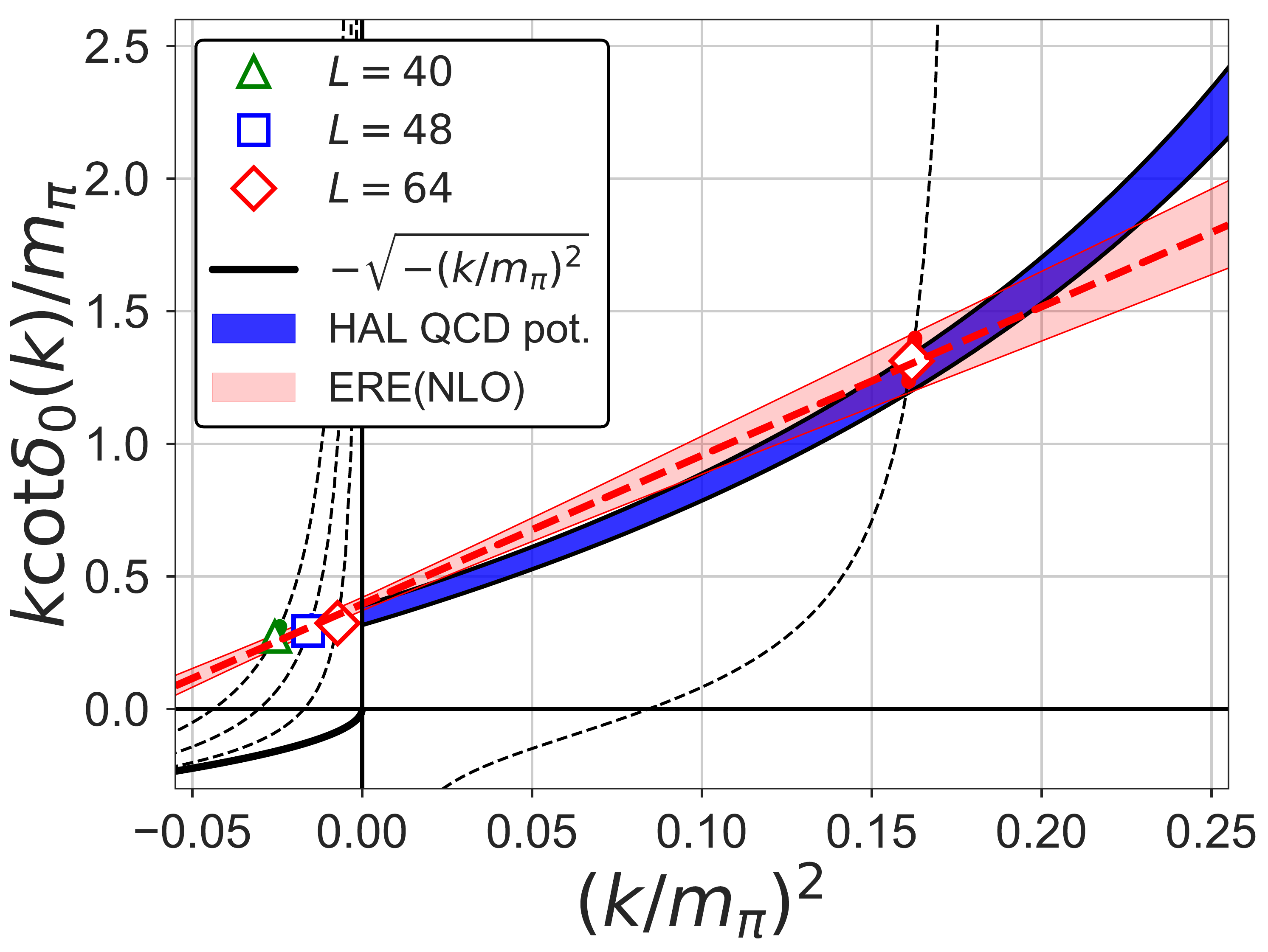}
  \caption{
    \label{fig:luscher}
    (Left) 
    The lowest eigenenergies on finite volumes from the HAL QCD potential.
      The red line corresponds to the fit by the asymptotic L\"uscher's finite volume formula in the large $L$, Eq.~(\ref{eq:L_dep_scat}).
    (Right) The scattering phase shifts from finite volume eigenenergies using L\"uscher's finite volume formula (green triangle, blue square, red diamonds),
    together with those in the infinite volume from the Schr\"odinger equation (blue band).
    The black dotted lines denote the constraints by the L\"uscher's finite volume formula,
    and the black solid line represents the bound state condition in the infinite volume.
    The red dashed line with the pink band corresponds to the NLO ERE analysis to the finite volume data.
  }
\end{figure}

Let us now calculate $k^2$ from  eigenvalue spectra of the Hamiltonian\footnote{
  Since this non-hermitian eigenvalue problem can be written 
  as the definite generalized Hermitian eigenvalue problem, eigenvalues are all real.
}
$H = H_0 + V_0^\mathrm{N^2LO}(r) + V_2^\mathrm{N^2LO}(r)\nabla^2$
on the finite volume ($L=40, 48, 64$) for the $A_1^+$ representation of the cubic group,
by employing fitted $V_0^\mathrm{N^2LO}(r)$ and $V_2^\mathrm{N^2LO}(r)$
at $L=64$ in Table~\ref{tab:fit_params}.
Fig.~\ref{fig:luscher}~(Left)
shows the volume dependence of the lowest eigenvalues:
The data are found to be well described by Eq.~(\ref{eq:L_dep_scat}), which 
indicates that the system does not have a bound state.
By fitting the data with Eq.~(\ref{eq:L_dep_scat}),
we obtain the scattering length
as $(a_0 m_\pi)^{-1} = 0.402(14)$ consistent with the values in Table~\ref{tab:scat_length}, 
$(a_0 m_\pi)^{-1} = 0.352(36){(^{+80}_{-0})}$.

As extensively discussed in Ref.~\cite{Iritani:2017rlk},
  the ERE, $k\cot\delta_0(k) = 1/a_0 + (1/2)r_\mathrm{eff}k^2 + \cdots$, 
  provides  a systematic and reliable way to relate the volume dependence of $\Delta E_L$,
  the scattering phase shifts and the bound state pole around $k^2=0$. 
\footnote{It was pointed out  in ~\cite{Iritani:2017rlk} that 
the  singular and/or unphysical behaviors of $k\cot\delta_0(k)$ around $k^2 =0$ 
can arise in the direct method ~\cite{Yamazaki:2015asa,Wagman:2017tmp, Berkowitz:2015eaa}  if the finite-volume
spectrum is not extracted reliably.}

In Fig.~\ref{fig:luscher}~(Right),
we plot the finite volume spectra
on the $(k^2, k\cot\delta_0(k))$ plane,
using the lowest eigenvalues of $H$  on $L = 40$, $48$, and $64$,
and the eigenvalue of the first excited state on $L = 64$.
Note that the data (triangle, square and diamonds) and their errors are
plotted together with the L\"uscher's formula (dotted lines).
The blue band corresponds to the results
obtained by solving the Schr\"odinger equation in the infinite volume.
We find that the   finite volume energy spectra at $k^2 < 0$ and $k^2 > 0$
are smoothly connected around $k^2=0$ along with the blue band,
as is expected from the analytic properties of $S$-matrix and the ERE.
In fact, the ERE at the NLO determined from these 4 data (pink band)
is consistent with the blue band at $|(k/m_\pi)^2| \lesssim 0.2$ within errors.
One also observes that
the positive intercept at $k^2=0$  ($1/a_0$) supports 
the conclusion from Fig.~\ref{fig:luscher}~(Left) that the system has no bound state.

\section{Summary}

In this paper, we have made critical investigations on
the systematic uncertainties in the HAL QCD method.
While the time-dependent HAL QCD method is
free from the issue associated with the ground state saturation,
the approximation of the energy-independent non-local potential
by the derivative expansion introduces systematic uncertainties, so that
it is necessary to check the errors introduced by the expansion.

We have performed the {(2+1)-flavor} lattice QCD calculation for the $\Xi\Xi(^1$S$_0)$ system
at $m_\pi = 0.51$~GeV.
Using the four-point correlation functions from both wall and smeared quark sources,
we have established the theoretical and numerical 
method to determine LO and N$^2$LO potentials
in the derivative expansion.
Scattering phase shifts calculated from these potentials reveal
that the LO potential is sufficient to reproduce observables at low energies
($k^2/m_{\pi}^2 < 0.1$), while
the N$^2$LO correction becomes non-negligible but
remains small even at high energies ($k^2/m_{\pi}^2 \simeq 0.5$), 
confirming the good convergence of the derivative expansion below
the inelastic threshold for this particular system.

We have also found that the potential at the LO analysis for  the wall source
agrees with the LO potential at  the N$^2$LO analysis except at short distances
and can reproduce the scattering phase shifts precisely at low energies.
Other systematic uncertainties such as  the inelastic state contaminations and
the finite volume effect to the potential are investigated
and are found to be well under control.

After establishing the reliability of the HAL QCD potential,
we have calculated the eigenvalues of the Hamiltonian in finite boxes with the potential.
The volume dependence of the lowest eigenvalues
is well described by $1/L$-expansion for scattering states obtained from the L\"uscher's finite volume formula.
We have also discussed the relation among the energy spectrum, phase shifts
and the effective range expansion.

In a forthcoming paper~\cite{Iritani:2018vfn},
  we will perform the spectral decomposition of the correlation function
  based on the eigenmodes of the Hamiltonian in a finite box with the HAL QCD potential,
 which will enable us to better understand the requirements for reliably extracting finite-volume energies.

\acknowledgments
We thank the authors of Ref.~\cite{Yamazaki:2012hi} and ILDG/JLDG~\cite{conf:ildg/jldg, Amagasa:2015zwb}
for providing the gauge configurations.
Lattice QCD codes of
CPS~\cite{CPS}, Bridge++~\cite{bridge++} and the modified version thereof by Dr. H.~Matsufuru, 
cuLGT~\cite{Schrock:2012fj} and domain-decomposed quark solver~\cite{Boku:2012zi,Teraki:2013}
are used in this study.
The numerical calculations have been performed on BlueGene/Q and SR16000 at KEK, HA-PACS at University of Tsukuba,
FX10 at the University of Tokyo and K computer at RIKEN, AICS (hp150085, hp160093).
This work is supported in part by the Japanese Grant-in-Aid for Scientific
Research (No. JP24740146, JP25287046, JP15K17667, JP16K05340, JP16H03978),
by MEXT Strategic Program for Innovative Research (SPIRE) Field 5,
by a priority issue (Elucidation of the fundamental laws and evolution of the universe) to be tackled by using Post “K” Computer,
and by Joint Institute for Computational Fundamental Science (JICFuS).

\clearpage

\appendix

\section{\label{app:N2LOcorr} {Non-locality vs. Energy dependence}}

Here we examine the relation between the energy-independent non-local potential
 with the derivative expansion,
   $U(\vec{r}, \vec{r'}) = \{V_0(r) + V_2(r) \nabla^2 + \cdots\}\delta(\vec{r} - \vec{r'})$,
    and the energy-dependent local potential, $V^{\rm eff} (r;E)$.
For simplicity, in this appendix,  we restrict ourselves to the N$^2$LO analysis. 
     In other words, we assume as if the non-local potential were given exactly by
   $U(\vec{r}, \vec{r'}) = \{V_0(r) + V_2(r) \nabla^2\} \delta(\vec{r} - \vec{r'})$.

In this case, it is easy to show that the Schr\"odinger equation with this non-local potential,
given by
\begin{equation}
  \left[ - \frac{\nabla^2}{2\mu} +
    V_0(r) + V_2(r)\nabla^2
      \right]\psi(\vec{r}) = E \psi(\vec{r}), \qquad \mu = m_B/2,
\end{equation}
can be written in terms of the energy-dependent local potential as
\begin{equation}
  \left[ - \frac{\nabla^2}{2\mu} + V^{\rm eff} (r;E)
         \right]\psi(\vec{r}) = E \psi(\vec{r}),
\end{equation}
where
\begin{equation}
  V^\mathrm{eff}(r; E) \equiv \frac{V_0(r) - m_B E V_2(r)}{1 - m_B V_2(r)},
  \label{eq:E_dep_pot}
\end{equation}
which gives an exact relation between the energy-independent non-local potential and  
the energy-dependent local potential (within the N$^2$LO analysis).
Although both descriptions for the potential are theoretically equivalent as shown above,  
we stress that the HAL QCD method is based on the energy-independent non-local potential,
which can be extracted from arbitrary linear combinations of the NBS wave function $\psi^W(\vec{r})$
thanks to the time dependent method, while the energy-dependent local potential requires the eigenstate saturation, which is difficult to achieve in practice, particularly for excited states.
Also $ V^\mathrm{eff}(r; E)$ gives the correct scattering phase shift at each $E$
(one potential per energy), while $V_0(r) + V_2(r)\nabla^2$ gives the correct scattering phase shifts (within the N$^2$LO analysis) at all $E \le E_{\rm th}$ (one potential for all).

\begin{figure}
  \centering
  \includegraphics[width=0.49\textwidth,clip]{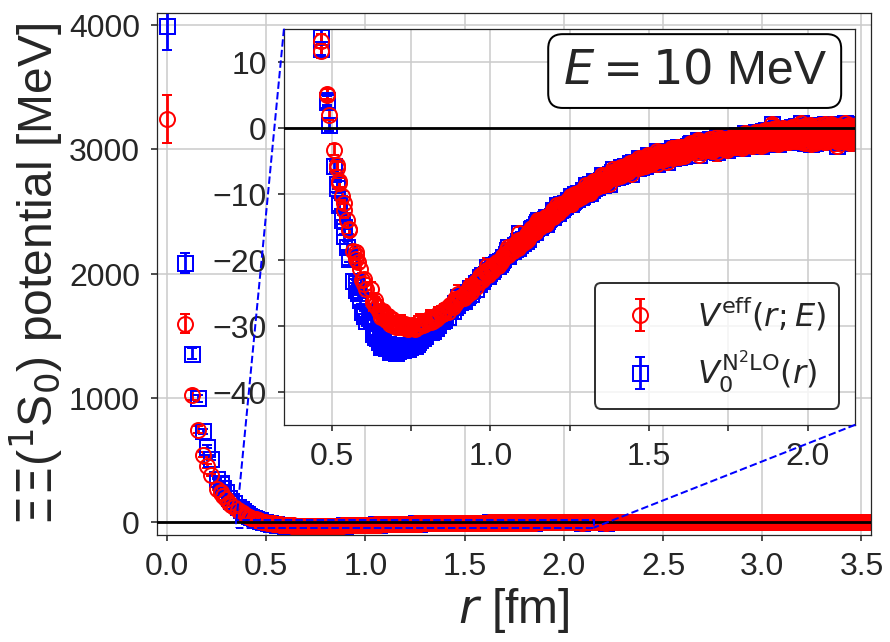}
  \includegraphics[width=0.49\textwidth,clip]{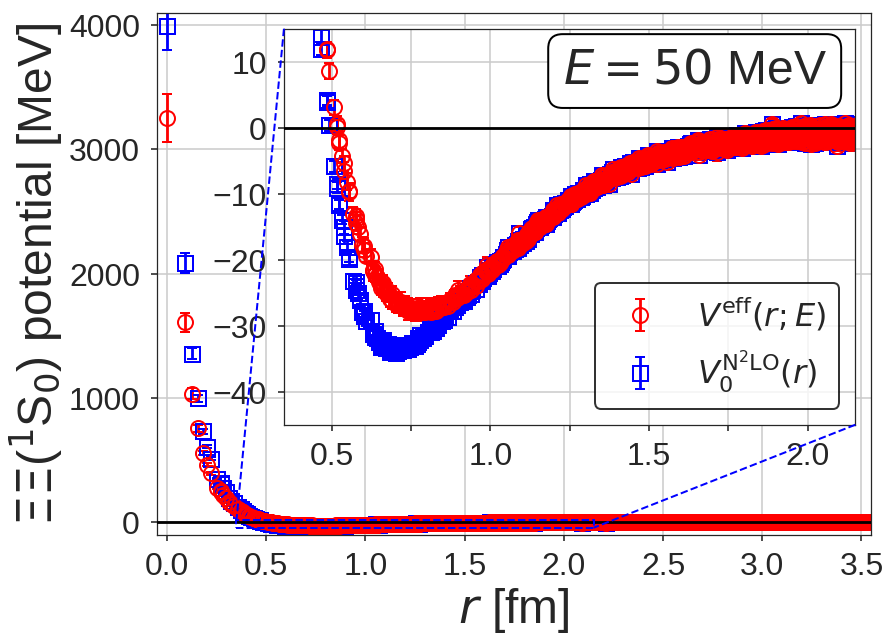}
  \includegraphics[width=0.49\textwidth,clip]{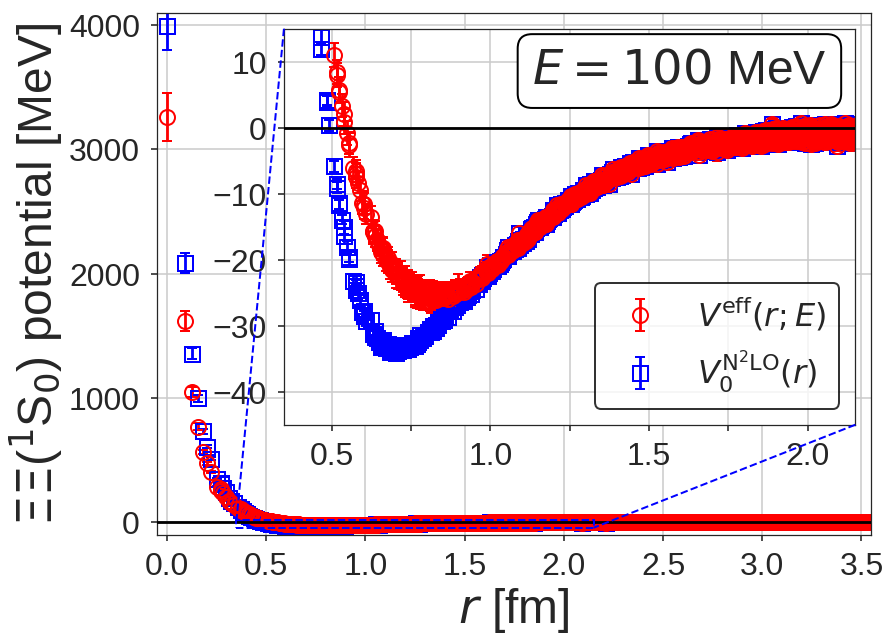}
  \includegraphics[width=0.49\textwidth,clip]{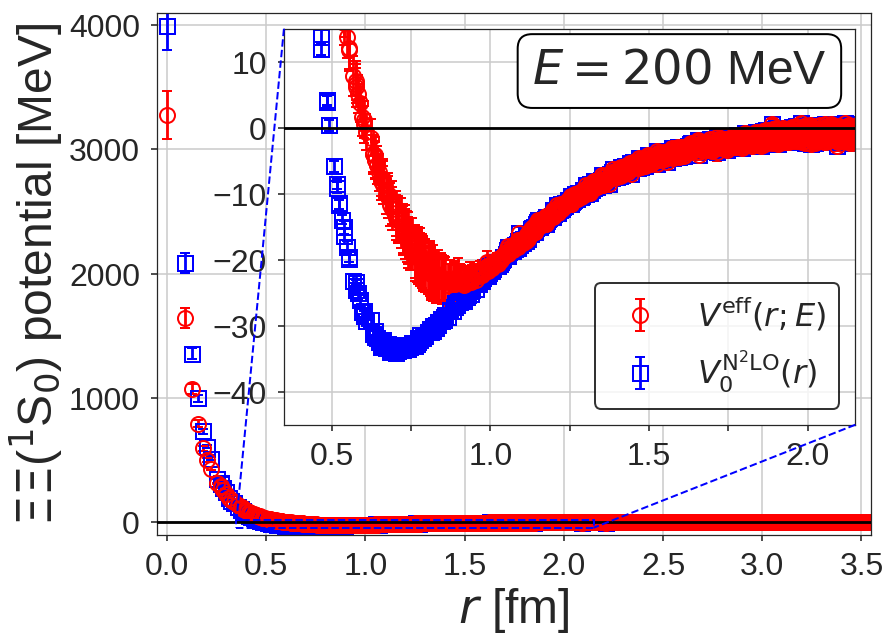}
  \caption{
    \label{fig:E_dep_pot}
    The energy dependence of the effective potential,
    $V^\mathrm{eff}(r;E)$, (red circles)
    compared with the LO potential at the N$^2$LO analysis, $V_0^\mathrm{N^2LO}(r)$ (blue  squares) at
     $E = 10$ MeV {(Top Left),} 
     $E = 50$ MeV {(Top Right),}
     $E = 100$ MeV {(Bottom Left) and} 
     $E = 200$ MeV {(Bottom Right)}. 
  }
\end{figure}

Fig.~\ref{fig:E_dep_pot} shows the energy dependence of  $V^\mathrm{eff}(r; E)$
from $E = 10$ MeV to $E = 200$ MeV.
 In these figures, we use $V_0^\mathrm{N^2LO}(r)$ and $V_2^\mathrm{N^2LO}(r)$ obtained at $t=13$ for $V_0(r)$ and  $V_2(r)$, respectively.
The energy dependent correction is small at low energies, while
it is no longer negligible at higher energies.
As the energy increases,
the attractive pocket at an intermediate distance becomes shallower
and the radius of the repulsive core becomes larger.

  This result also demonstrates how the non-locality of the energy-independent potential,
  $U(r,r')$,
  (Note that $V_0(r)$, $V_2(r), \cdots$ are energy-independent by definition
\footnote{ In the literature, there appears a confusion on the relation between
    the energy-independent non-local potential and the energy-dependent local potential.
    See Ref.~\cite{Aoki:2017yru}, which clarifies the relation between the two in detail.
}),
  is related to the energy dependence of the local potential, $V^\mathrm{eff}(r; E)$.

\end{document}